\documentclass[]{aa}  
\usepackage{amsmath}
\usepackage{bm}
\usepackage[colorlinks,breaklinks]{hyperref}
\hypersetup{linkcolor=blue,citecolor=blue,filecolor=black,urlcolor=blue}

\usepackage{amssymb}

\newcommand\uncIa{`Ia-unclear'}

\newcommand{\kms}{km s$^{-1}$}

\newcommand{\MgII}{Mg~{\sc ii} }

\newcommand{\OI} {\ion{O}{i}}
\newcommand{\SiII} {\ion{Si}{ii}}
\newcommand{\SiIIs} {\SiII\ $\lambda6355$}
\newcommand{\SiIIf} {\SiII\ $\lambda5972$}
\newcommand{\SiIId} {\SiII\ $\lambda4130$}
\newcommand{\CaII} {\ion{Ca}{ii} }
\newcommand{\SII} {\ion{S}{ii} }
\newcommand{\SIIone} {\SII\ $\lambda5454$}
\newcommand{\SIItwo} {\SII\ $\lambda5640$}

\newcommand{\TiII} {\ion{Ti}{ii} }
\newcommand{\Nifs}{$^{56} \rm Ni$}
\newcommand{\Msun}{\ensuremath{{M}_\odot}}
\newcommand{\Mstar}{\ensuremath{{M}_\star}}

\newcommand{\md}[1]{{\color{red} #1}}

\newcommand{\angstrom}{\mbox{\normalfont\AA}}
\usepackage{xspace}

\usepackage{graphicx}

\usepackage{txfonts}
\usepackage{pdflscape}
\usepackage{comment}
\usepackage{orcidlink}
\usepackage{hyperref}

\begin{document}

%
%
    \title{ZTF SN~Ia DR2: The spectral diversity of Type Ia supernovae in a volume-limited sample}
    \author{U. Burgaz \inst{1}\orcidlink{0000-0003-0126-3999}
    \and K. Maguire\inst{1}\orcidlink{0000-0002-9770-3508}
    \and G. Dimitriadis\inst{1}\orcidlink{0000-0001-9494-179X}
    \and L. Harvey\inst{1}\orcidlink{0000-0003-3393-9383}
    \and R. Senzel\inst{1}\orcidlink{ 0009-0002-0243-8199}
    \and J. Sollerman\inst{2}\orcidlink{0000-0003-1546-6615}
    \and J. Nordin\inst{3}\orcidlink{0000-0001-8342-6274}
    \and L.~Galbany\inst{4,5}\orcidlink{0000-0002-1296-6887}
    \and M.~Rigault\inst{6}\orcidlink{0000-0002-8121-2560}
    \and M.~Smith\inst{7}\orcidlink{0000-0002-3321-1432}
    \and A.~Goobar\inst{8}\orcidlink{0000-0002-4163-4996}
    \and J.~Johansson\inst{8}\orcidlink{0000-0001-5975-290X}
    \and P.~Rosnet\inst{9}    
    \and M.~Amenouche\inst{10}\orcidlink{0009-0006-7454-3579}
    \and M.~Deckers\inst{1}\orcidlink{0000-0001-8857-9843}
    \and S.~Dhawan\inst{11}\orcidlink{0000-0002-2376-6979}
    \and M.~Ginolin\inst{6}\orcidlink{0009-0004-5311-9301}
    \and Y.-L.~Kim\inst{7}\orcidlink{0000-0002-1031-0796}
    \and A.~A.~Miller\inst{12,13}\orcidlink{0000-0001-9515-478X}
    \and T.~E.~Muller-Bravo\inst{4,5}\orcidlink{0000-0003-3939-7167}
    \and P.~E.~Nugent\inst{14,15}\orcidlink{000-0002-3389-0586}
    \and J.~H.~Terwel\inst{1,16}\orcidlink{0000-0001-9834-3439}
    \and R.~Dekany\inst{17}\orcidlink{0000-0002-5884-7867}
    \and A.~Drake \inst{18}
    \and M.~J.~Graham\inst{19}\orcidlink{0000-0002-3168-0139}
    \and S.~L.~Groom\inst{14}\orcidlink{0000-0001-5668-3507}
    \and M.~M.~Kasliwal\inst{19}\orcidlink{0000-0002-5619-4938}
    \and S.~R.~Kulkarni\inst{19}
    \and K.~Nolan\inst{19}\orcidlink{0009-0002-4724-7118}
    \and G.~Nir\inst{14}\orcidlink{0000-0002-7501-5579}
    \and R.~L.~Riddle\inst{17}\orcidlink{0000-0002-0387-370X}
    \and B.~Rusholme\inst{20}\orcidlink{0000-0001-7648-4142}
    \and Y.~Sharma\inst{19}\orcidlink{0000-0003-4531-1745}
    }

   \institute{School of Physics, Trinity College Dublin, College Green, Dublin 2, Ireland\\
            \email{burgazu@tcd.ie}
            \and Oskar Klein Centre, Department of Astronomy, Stockholm University, SE-10691 Stockholm, Sweden\
            \and Institut für Physik, Humboldt-Universit\"at zu Berlin, Newtonstr. 15, 12489 Berlin, Germany\
            \and Institute of Space Sciences (ICE, CSIC), Campus UAB, Carrer de Can Magrans, s/n, E-08193, Barcelona, Spain\   
            \and Institut d'Estudis Espacials de Catalunya (IEEC), E-08034 Barcelona, Spain\  
            \and Univ Lyon, Univ Claude Bernard Lyon 1, CNRS, IP2I Lyon/IN2P3, UMR 5822, F-69622, Villeurbanne, France\ 
            \and Department of Physics, Lancaster University, Lancs LA1 4YB, UK\
            \and Oskar Klein Centre, Department of Physics, Stockholm University, SE-10691 Stockholm, Sweden\
            \and Universit\'e Clermont Auvergne, CNRS/IN2P3, LPCA, F-63000 Clermont-Ferrand, France\            
            \and National Research Council of Canada, Herzberg Astronomy \& Astrophysics Research Centre, 5071 West Saanich Road, Victoria, BC V9E 2E7, Canada\
            \and Institute of Astronomy and Kavli Institute for Cosmology, University of Cambridge, Madingley Road, Cambridge CB3 0HA, UK\
            \and Department of Physics and Astronomy, Northwestern University, 2145 Sheridan Rd, Evanston, IL 60208, USA\
            \and Center for Interdisciplinary Exploration and Research in Astrophysics (CIERA), Northwestern University, 1800 Sherman Ave, Evanston, IL 60201, USA\
            \and Lawrence Berkeley National Laboratory, 1 Cyclotron Road, MS 50B-4206, Berkeley, CA 94720, USA\
            \and Department of Astronomy, University of California, Berkeley, 501 Campbell Hall, Berkeley, CA 94720, USA\  
            \and Nordic Optical Telescope, Rambla Jos\'e Ana Fern\'andez P\'erez 7, ES-38711 Bre\~na Baja, Spain\  
            \and Caltech Optical Observatories, California Institute of Technology, Pasadena, CA, 91125, USA\
            \and Cahill Center for Astrophysics, California Institute of Technology, MC 249-17, 1200 E California Boulevard, Pasadena, CA, 91125, USA\
            \and Division of Physics, Mathematics and Astronomy, California Institute of Technology, Pasadena, CA, 91125, USA\
            \and IPAC, California Institute of Technology, 1200 E. California Boulevard, Pasadena, CA, 91125, USA\
}

\titlerunning{ZTF SN Ia DR2 - Spectral Diversity of SNe Ia}
\authorrunning{U. Burgaz, et al.}

\date{}
 
  \abstract
  {More than 3000 spectroscopically confirmed Type Ia supernovae (SNe Ia) are presented in the Zwicky Transient Facility SN Ia Data Release 2 (ZTF DR2). In this paper, we detail the spectral properties of 482 SNe Ia near maximum light, up to a redshift limit of \textit{z} $\leq$ 0.06. We measure the velocities and pseudo-equivalent widths (pEW) of key spectral features (\SiIIf\ and \SiIIs) and investigate the relation between the properties of the spectral features and the photometric properties from the SALT2 light-curve parameters as a function of spectroscopic sub-class. We discuss the non-negligible impact of host galaxy contamination on SN Ia spectral classifications, as well as investigate the accuracy of spectral template matching of the ZTF DR2 sample. We define a new subclass of underluminous SNe Ia (`04gs-like') that lie spectroscopically between normal SNe Ia and transitional 86G-like SNe Ia (stronger \SiIIf\ than normal SNe Ia but significantly weaker \TiII\ features than `86G-like' SNe).   We model these `04gs-like' SN Ia spectra using the radiative-transfer spectral synthesis code \textsc{tardis} and show that cooler temperatures alone are unable to explain their spectra; some changes in elemental abundances are also required. However, the broad continuity in spectral properties seen from bright (`91T-like') to faint normal SN Ia, including the transitional and 91bg-like SNe Ia, suggests that variations within a single explosion model may be able to explain their behaviour.}


   \maketitle

\section{Introduction}

Type Ia supernovae (SNe Ia) are  commonly accepted to be the thermonuclear explosion of a carbon–oxygen (C/O) white dwarf (WD) in a binary system \citep{hoyle1960, maoz2014}. SNe Ia are proven to be reliable distance estimators and are used as “standardisable candles” to measure extragalactic distances, where observations of distant SNe Ia led to the discovery of the accelerating expansion of the Universe \citep{perlmutter1997, perlmutter1999, ries1998, ries2016}. However, the exact nature and the explosion mechanism of their progenitors still remain in question \citep[e.g.,][]{khokhlov1991, maeda2016}. Even though several progenitor scenarios are proposed for SNe Ia, the most commonly considered progenitor scenarios are the single degenerate scenario  \citep[SD, ][]{whelan1973, nomoto1997}, which consists of a C/O WD and a non-degenerate companion such as a main-sequence star or an evolved star (Helium star or a red giant), and the double degenerate model \citep[DD, ][]{iben1984, webbink1984}, which consists of two WDs that result in the explosion of one or both of the WDs. 

The spectra of SNe Ia around maximum light are relatively homogeneous, with absorption lines of intermediate-mass elements dominating, such as \SiIIs, \SiIIf, \SiIId, \CaII H\&K, \CaII near-infrared triplet, \SIIone\ and \SIItwo\ \citep{zhao2021}. While \SiIIs\ absorption is the most prominent feature in SNe Ia spectra near maximum light, in peculiar cases such as SN 1991T \citep{filippenko1992b, phillips1992} and SN 1999aa \citep{filippenko1999} this feature appears extremely weak. These SNe Ia (91T-like, 99aa-like) exhibit slow decline rates in their light curves and have higher peak absolute magnitudes on average. Strong \TiII absorption around 4300 \AA\ is also present in some subluminous classes of SNe Ia \citep[e.g., SN 1991bg-like;][]{filippenko1992a}. The estimated brightness of 91bg-like SNe Ia is roughly two magnitudes lower than that of a normal SN Ia. `Transitional' events also exist that have spectral features and luminosities between normal and faint SNe Ia, such as SN 1986G \citep{phillips1987} that displayed weaker \TiII absorption than seen in 91bg-like SNe Ia around maximum light.  The variation in spectral features in maximum light SN Ia spectra is most likely due to varying temperatures in the ejecta \citep{nugent1995, hachinger2008} but the connection between luminosity, temperature, and decline rate is still uncertain.

Several classification schemes have been proposed to investigate the spectroscopic diversity seen in SNe Ia, in particular the velocity evolution with time of the \SiIIs\ feature, as well as the relative strength of the \SiIIs\ and \SiIIf\ features. Based on the velocity evolution of \SiIIs, \citet{benetti2005} divided SNe Ia into three groups: `high-velocity gradient (HVG)', `low-velocity gradient (LVG)', and a `faint' group that consisted of subluminous 91bg-like events. Another classification scheme based on the pseudo equivalent width (pEW) of the \SiII\ absorption features of 5972 and 6355 \angstrom\ was introduced by \cite{branch2006, branch2009}. There are four subgroups in this scheme: `core normal', `cool', `broad line', and `shallow silicon'. \citet{wang2009_may_a} proposed a classification scheme purely based on the velocity of the \SiIIs\ absorption feature near maximum light, where SNe Ia with expansion velocities, $v_{\mathrm{Si}}$, of $v_{0} \gtrsim$ ~12,000 km s$^{-1}$ are defined as `High Velocity (HV) SNe Ia' and those with $v_{\mathrm{Si}} \lesssim$ ~12,000 km s$^{-1}$ are defined as `Normal Velocity (NV) SNe Ia'. The observed diversity in velocity is suggested to be independent of light curve evolution and may be attributed, at least in part, to a viewing angle effect in explaining the diversity of HV/NV SNe Ia \citep{maeda2010}.  It has been suggested that HV SN Ia exhibit differences in their light curve evolution  \citep{wang2008, wang2009_may_a, burgaz2021}, shorter explosion to maximum light rise time \citep{ganeshalingam2011}, have redder intrinsic \textit{B-V} colour \citep{pignata2008}, may have different extinction laws \citep{wang2009_may_a, foleykasen2011} and may be more concentrated towards the centres of their host galaxies \citep{wang2013}. 

In this paper, we present the analysis of the spectroscopic features of the second Zwicky Transient Facility (ZTF) Data Release (DR2) SNe Ia, obtained as part of the Zwicky Transient Facility survey from 2018 - 2020 \citep[][]{bellm2019, graham2019, masci2019, dekany2020}, as well as the link to their photometric properties. The ZTF is an optical time-domain survey that  has been in operation since 2018. It spectroscopically confirmed nearly 4000 low-redshift SNe Ia between 2018 and 2020, without biases in terms of host-galaxy properties (i.e. massive galaxies were not specifically targeted). In Section~\ref{sec:dr2intro1}, we present the sample selection, spectroscopic classifications and the data reduction. Section~\ref{sec:analysis} discusses the measurements of spectral features (velocities and equivalent widths) and the accuracy of spectral template matching SN Ia sub-classes. In Section~\ref{sec:results}, we examine the spectroscopic sub-classes and how photometric properties and spectral features are related. Conclusions and possible future works are presented in Section~\ref{sec:conclusions}.

\section{Sample selection and spectral measurements}
\label{sec:dr2intro1}
In this section, we discuss the overall properties of the ZTF SNe Ia in the DR2 and the methods used to analyse their spectra. In Section \ref{sec:sampleselect}, we describe the ZTF SN Ia sample, including redshift determination, light-curve parameters, optical spectroscopy, and the methodology for their initial spectral classifications.  In Section \ref{sec:sampleselect}, we detail the cuts made on the sample to obtain our desired maximum light spectral sample. Finally, in Section \ref{sec:sec2velocities}, we present the method used to measure properties, such as velocities and equivalent widths, from the spectra.

\subsection{ZTF SN Ia Data Release 2}
\label{sec:dr2intro}

The ZTF DR2 SN Ia sample contains 3628 spectroscopically confirmed SNe Ia discovered from 2018 to 2020 (see \md{Rigault et al.~subm.} and \md{Smith et al.~in prep.}~ for further details) that had optical ZTF light curves obtained for them. Thanks to the Bright Transient Survey (BTS) program, most of them were obtained near or before peak brightness. 61\% of the ZTF SN Ia DR2 sample has host spectroscopic redshifts, among which 71\% coming from DESI through the MOST Host program \citep{MOST_Hosts}. 9\% comes from identifiable host galaxy lines in the SN spectra and the remaining 30\% are derived from template matching using \textsc{snid} \citep{Blondin2007}

\md{Smith et al.~(in prep.)} compared the known redshift values from host catalogues and \textsc{snid} redshifts for SNe Ia in the sample and found good overall agreement with a redshift dispersion of $3 \times 10^{-3}$, corresponding to a velocity dispersion of $\sim500$ \kms. 

The majority of optical spectra from ZTF come from the low-resolution integral field spectrograph, SED machine \cite[SEDm,][]{blagorodnova2018, rigault2019, kim2022} on the 60-inch telescope at the Mount Palomar observatory. These spectra are supplemented by spectra from several other telescopes with higher resolution, as discussed in \md{Smith et al. (in prep.)}. All the spectra are corrected to the rest-frame using an estimate of its host galaxy redshift and corrected for Milky Way extinction \citep{schlafly2011} using a total-to-selective extinction ratio, R$_V$ of 3.1.

As part of the ZTF DR2 analysis, a spectral typing effort was made to identify the sub-classifications of the DR2 sample into the classes of `normal SN Ia' \citep{wang2009_jul_b, nugent2011}, `91bg-like' \citep{filippenko1992a}, `91T-like'\citep{filippenko1992b}, `Ia-peculiar' or if the sub-classification was unclear, `SN Ia-unclear'. The classification were accomplished in two steps. Firstly, users were invited to a specially designed web tool called the `typingapp'\footnote{\url{https://typingapp.in2p3.fr/dashboard}} that allowed them to see best \textsc{SNID} fits of each individual spectra with several sub-types and the phase-redshift distributions. These \textsc{snid} fits \citep{Blondin2007} using priors based on the light-curve phase estimates from the SALT2 \citep{guy2007} light curve fit  and a redshift prior based on host galaxy spectral features, where this was available. Its sole purpose is to classify all SNe with spectral information only (hence, no host information was shown to users). Users could see the light curve but this was only for confirming spectral phases relative to the light curve phase. Then, each user selected and reported sub-types for the SNe on the typingapp. The spectra of each SN Ia, along with their best fitting \textsc{snid} matches, were visually inspected by multiple members of the ZTF SN Ia working group to determine the most likely sub-classification for each event. Secondly, a more thorough subclassification, targeted at the events with $z\leq0.06$ (the redshift range studied in this work), was performed (see Section \ref{sec:template_matching}). A detailed methodology for these classifications is also presented in \md{Dimitriadis et al.~(in prep.)}.

\subsection{Sample Selection}
\label{sec:sampleselect}

\begin{table}
\centering
\caption{Summary of the cuts applied to obtain our final spectral sample.}
\label{tab:cuts}
\begin{tabular}{l c c}
\hline\\[-0.5em]
     & Criteria & No. of SNe \\[0.15em]
    \hline\\[-0.8em]
    \hline\\[-0.5em]
Full ZTF SN Ia DR2 & & 3628\\[0.30em]
with good LC coverage & & 2959\\[0.30em]
with basic LC quality cut*&  & 2790\\[0.30em]
with redshift cut & $z$ $\leq$ 0.06 & 1093\\[0.30em]
with phase cut & $-$5 d $\leq$ $t_0$ $\leq$ 5 d & 546\\[0.30em]
with non-peculiar SNe Ia & & 526\\[0.30em]
with spectra quality cut & SNR $\geq$ 5 & 482\\[0.30em]
    \hline\\[-0.5em]  
\end{tabular}
 \begin{flushleft}
*Cuts of $-4 < x_1<4$, $\sigma_{x_1} < 0.6$, $c < 0.8$, $\sigma_{c} < 0.35$, $\sigma_{t_0} < 1$ d (see Section \ref{sec:sampleselect}).
 \end{flushleft}

\end{table}

\begin{figure*}
\centering
\includegraphics[width=\textwidth]{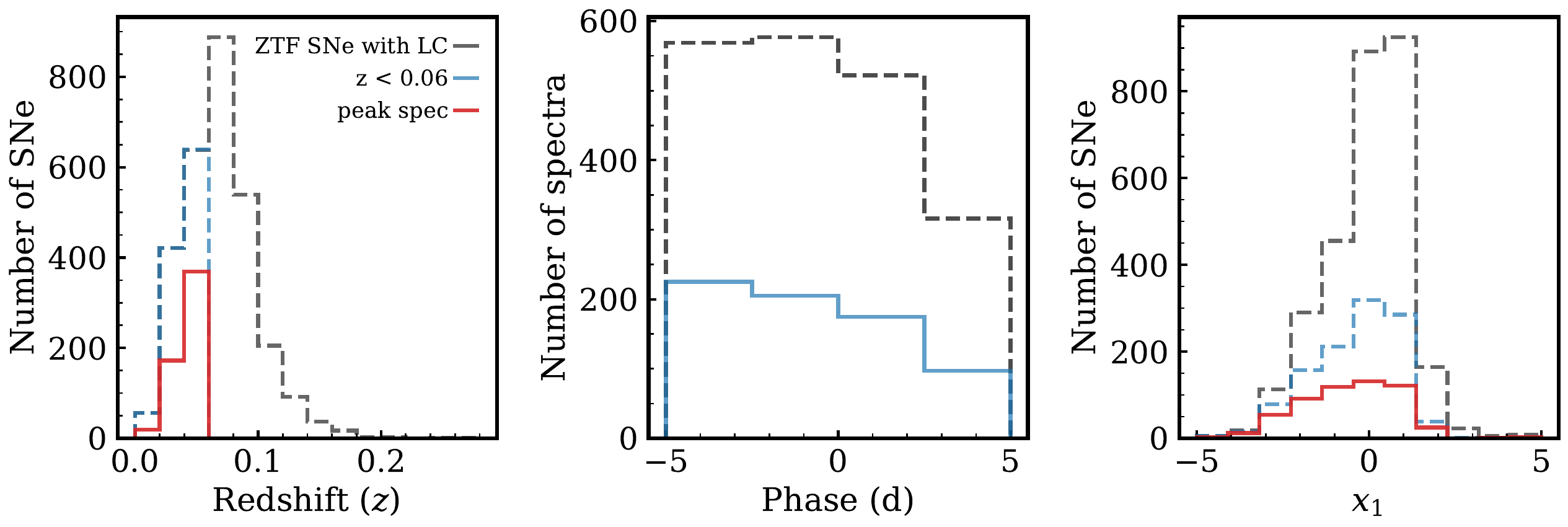}
\caption{\textbf{Left}: Redshift distribution of the full ZTF SN Ia sample (grey dashed line, 2959 SNe Ia) with good light curve coverage, the sample with a redshift limit at \textit{z} $\leq$ 0.06 (blue dashed line, 1093 SNe Ia), and SNe Ia passing our other cuts that have a spectrum at $-$5 to 5 d with respect to maximum light (red solid line, 482 SNe Ia). \textbf{Middle}: The rest-frame phase distribution of all good light curve coverage ZTF SNe Ia spectra at $-$5 to 5 d (grey) and with the redshift cut of \textit{z} $\leq$ 0.06. \textbf{Right}: SALT2 $x_1$ distribution of the ZTF SN Ia sample.}
\label{fig:hist}
\end{figure*}

In this study, we focus on an analysis of maximum light spectra of SNe Ia and therefore, apply a number of cuts on the photometry and spectra to achieve a relatively clean and unbiased sample with well constrained spectral phases. Table~\ref{tab:cuts} shows all the criteria used to define the final sample along with the number of remaining SNe Ia after each cut. 

The ZTF \textit{gri}-band light curves of the SNe Ia in the sample were fit with version 4\footnote{\url{https://sncosmo.readthedocs.io/en/stable/}} of the SALT2 light-curve fitter \citep{guy2007}, which gives the light-curve stretch, \textit{x$_1$}, colour $c$, and the time of maximum light, \textit{t$_0$} \md{(Rigault et al.~subm.)}. To select a suitable sample for this work with reliable phases and light curve properties to have enough photometry data, we follow the light curve coverage constraints of \md{Smith et al.~(in prep.)} of requiring each SN to have at least two detections in two filters both pre- and post-maximum light (relative to \textit{t$_0$}) and a total of at least seven detections across all filters. In order to have a well constrained sample, we further applied the following criteria: \textit{x$_1$} of $-$4 to 4 with an uncertainty $<0.6$; $c$ < 0.8 with an uncertainty $<0.35$.

Based on survey simulations, the ZTF DR2 sample is expected to be complete for normal SNe Ia up to a redshift (\textit{z}) of $z$ $\leq$ 0.06 \md{(Amenouche et al.~in prep.)}. After the photometry cuts, we select SNe Ia with $z \leq$ 0.06 and that have at least one spectrum between $-$5 to 5 days with respect to the  time of maximum light ($t_0$) with an uncertainty of $<1$d, where the velocity evolution within this range has been shown to be small \cite[e.g.][]{maguire2014}. 

We have included the bright 91T-like, bright transitional 99aa-like, normal, transitional \citep{Hsiao2015, Gall2018} and 91bg-like objects in our sample. We disregarded all the peculiar subtypes. These subtypes include; `03fg-like' SNe Ia \citep{Howell2006}, which are overluminous and show slow evolving light-curve features, `02cx-like' SNe Ia \citep{li2003} that are underluminous and spectroscopically weird, `02es-like' SNe Ia \citep{ganeshalingam2012} that look similar to underluminous SNe Ia with a slow light-curve evolution, `SN Ia-csm' \citep{Benetti2006} that show strong interaction with circumstellar medium around them \citep{Dilday2012}. These peculiar objects and their properties are explained and studied in detail in \md{Dimitriadis et al.~(in prep.)}

The initial spectral typing (\md{Dimitriadis et al.~in prep.}) for our sample of 482 SNe Ia resulted in preliminary identification of 377 normal SNe Ia, 35 91bg-like, 18 91T-like, and 52 SNe Ia (those without sub-classification, which will be referred to as \uncIa\ throughout the text). However, there is a subjective aspect to some of these classifications, e.g., where no further sub-classifications are taken into account while typing and making a final decision. One of our main aims was to assess the accuracy of this spectral template matching process and assign sub-classes using objective measurements of features present in the spectra, and this is discussed further in Section \ref{sec:template_matching}.

In order to measure the velocities and pseudo-equivalent widths (pEW) from the \SiIIs\ feature, we only selected the spectra with high signal-to-noise ratio (SNR) higher than 5. The SNR is estimated from the \SiIIs\ absorption feature, where the estimate is the ratio of the depth of the feature to the standard deviation in the selected continuum region.  After all the cuts were applied, the final sample consists of 657 spectra of 482 unique SNe Ia between -5 to 5 days (Table~\ref{tab:cuts}). The redshift, phase and light curve width parameter $x_1$ distributions are presented in Fig.~\ref{fig:hist}, split into the good light curve coverage ZTF DR2 SN Ia sample (grey), the sample after the redshift cut of $z \leq$ 0.06 has been applied (blue), and the maximum light ($-$5 to 5 days) sample (red).

\subsection{Velocity and width measurements}
\label{sec:sec2velocities}

\begin{figure*}
\centering
\includegraphics[width=\textwidth]{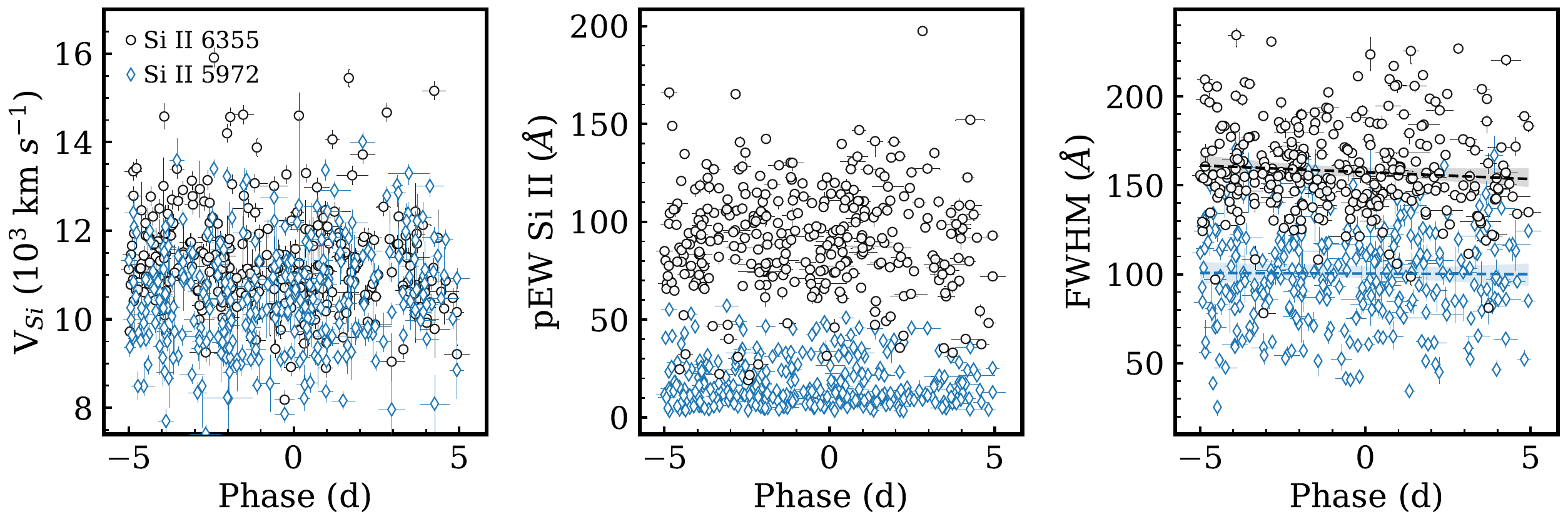}
\caption{The velocity (left), pEW (middle) and FWHM (right) of \SiIIf\ (blue diamonds) and \SiIIs\ (black circles) from the ZTF DR2 SN Ia sample as a function of rest-frame phases with respect to SALT2 maximum light. The black and blue dashed lines in the right panel represent the linear regression fits to the FWHM of \SiIIs\ and \SiIIf, respectively.}
\label{fig:velpEWdist}
\end{figure*}

In order to study the spectroscopic properties of our SN Ia sample, we have performed several measurements on different spectral features.  There are several methods to fit SN Ia spectral features seen in the optical spectra. \citet{blondin2006} used a method based on smoothing the spectra and choosing the minimum value to obtain the line velocities.  \citet{childress2013} used a different method by applying multiple Gaussian fittings since there are triplets such as \CaII NIR that overlap.  In this work, we adapt the method of \citet{childress2013} to obtain the line expansion velocities and pseudo-equivalent widths of \SiIIf\ and \SiIIs. 

We start our study of the spectral absorption features by first defining a continuum for each feature. This is done visually for each spectrum by selecting regions on either side of the feature to which a line is fitted to define the continuum. The spectrum is then divided by the continuum to normalize each absorption profile. We apply a double Gaussian fit for the absorption features of \SiIIs\ line, which is a doublet of \SiII\ $\lambda$6347 and $\lambda$6371 features. We force both doublet components to have the same widths,  velocities, and strengths since these lines tend to saturate in the assumed optically thick regime \citep{childress2013}. Another double Gaussian fit is applied to the absorption features of \SiIIf\ doublet (\SiII\ $\lambda$5958 and $\lambda$5979), where we again fix the individual line components to have same widths, velocities and strengths.

We estimate the uncertainties on the fit parameters by varying the positions of the visually picked continuum regions on other side of the feature and repeating the Gaussian fitting. These uncertainties are combined with the uncertainties on the individual fits to give a total uncertainty for the fit parameters. We do not want to exclude the SNe Ia with template-matched redshifts because this may bias our sample by preferentially removing fainter host galaxies that are are less likely to have a host galaxy spectrum. For 105 SNe Ia in our sample, the redshift comes from \textsc{snid} template matching, and in order to account for the redshift uncertainty discussed in Section \ref{sec:dr2intro}, we add 500 \kms\ in quadrature to the \SiII\ velocity uncertainties. 

The pseudo-equivalent width (pEW) of each feature is calculated based on the Gaussian fit parameters.  For some SNe Ia in the sample, the \SiIIf\ feature is quite weak and for these events, we calculated the 3$\sigma$ upper limit of the \SiIIf\ feature by forcing a fit at the expected \SiIIf\ location to find the residuals of the fit and calculate the root mean square of the residuals. We infer the full width at half maximum (FWHM) of the \SiIIf\ feature based on the relation between the FWHM of \SiIIs\ and of \SiIIf\ for SNe Ia where both exist. For that, we first calculated the mean FWHM of \SiIIf\ (159$\pm$25 \AA) and \SiIIs\ (100$\pm$28 \AA), where both exist, to estimate a scaling factor of $\sim$ 1.6 (see the right panel of Fig. \ref{fig:velpEWdist} for the two fits, each showing similar linear relations that yields a scaling factor of $\sim$ 1.6). After correcting for the relation, we use the estimated FWHM of \SiIIf\ to calculate the 3$\sigma$ upper limits of its pEW.

The velocity, pEW and FWHM of the \SiIIf\ and \SiIIs\ features are shown as a function of phase in Fig.~\ref{fig:velpEWdist}. The results of the fits for each feature and spectrum are given in Table~\ref{tab:spec_p}. The distributions of velocity, pEW and FWHM are seen to be relatively flat in the phase range studied here of $-$5 to +5 d with respect to maximum light. No further correction for evolution within this phase range is applied.

\section{Analysis}
\label{sec:analysis}

One of the aims of our study is to assess the accuracy of the sub-classification of the DR2 SN Ia sample that used the method explained above (Section \ref{sec:dr2intro}). We also investigate the possible spectral transition from the `normal' to `91T-like' SNe Ia in Section \ref{sec:normal_to_91T} and `normal' to `91bg-like' SNe Ia in Section \ref{sec:normal_to_91bg}. The impact of host galaxy contamination on SN Ia classifications and the \SiII\ feature is discussed in Section \ref{sec:host_contamination} and Section \ref{sec:host_contamination_SiII}, respectively.

\subsection{Spectral (re-)classification of the sample}
\label{sec:template_matching}

\begin{table*}
\centering
\caption{Sub-type reclassification in ZTF DR2. The initial column is the original sub-classification from the DR2 typing effort. The numbers below do not represent the final numbers of each sub-type but only indicate changes in sub-typing from the initial assessment to our investigation. The number of events in final sample after removal of host contaminated events is given in Table \ref{tab:DLR_cut_table}.}
\label{tab:reclass}
\begin{tabular}{lccccccccc}

\hline\\[-0.5em]
Sub-type & Initial & 91T-like & 99aa-like & normal &04gs-like & 86G-like & 91bg-like & Ia-unclear \\
    \hline\\
91T-like & 18 & - & 9 & 2 & - & - & - &1\\
normal & 377 & - & 18 & - & 18 & 3 & - & 55\\
91bg-like & 35 & - & - & - & - & 8 & - & 3\\
Ia-unclear & 52 & 1 & 12 & 10 & - & - & - & -\\
\hline\\
\end{tabular}
\end{table*}

To determine the accuracy of spectral-template matching, we have inspected in detail the sub-classifications from the ZTF DR2 initial analysis of all 482 SNe Ia in our maximum light sample. As discussed previously, we do not study the `Ia peculiar' sub-type in this analysis. In order to check the accuracy of the initial DR2 sub-typing, we visually inspected each SN Ia and compared to known examples of each sub-class that we wanted to investigate. We created a comparison sample from the peak spectra of SN 1991T for `91T-like' SNe, SN 1999aa for bright transitional/`99a-like' SNe, SN 2011fe for normal SNe Ia, SN 2004gs for faint transitional/`04gs-like' SNe, SN 1986G for `86G-like' SNe and SN 1991bg for `91bg-like' SNe. All spectra in our sample were visually inspected and over-plotted against different subtypes to identify the closest match for each, taking into account all key features. In Table \ref{tab:reclass}, we show the initial classifications of the DR2, as well as our changes to those initial categories.

We are particularly interested in constraining the diversity of the events at the two transition boundaries $-$ from normal to overluminous (e.g., `91T-like') and from normal to subluminous (e.g., `91bg-like'). Therefore, we are investigating subtle differences in their spectral properties.  In Section \ref{sec:normal_to_91T}, we discuss the sub-classification criteria for SNe Ia around the normal to overluminous boundary (`99aa-like, 91T-like'). In Section \ref{sec:normal_to_91bg}, we discuss the SN Ia sub-classification at the normal to underluminous boundary (`04gs-like', `86G-like', `91bg-like').

There are 52 SNe Ia that were initially classified (prior to our analysis) as `Ia-unclear', meaning that a sub-classification had not been made. In this work, we have reclassified 23 of these cases into sub-classifications: `91T-like', `99aa-like', and `normal'. None of the `Ia-unclear' were reclassified into the underluminous sub-classes. The initial DR2 classification exercise identified 377 normal SNe Ia. However, we updated these classifications to find a total 295 events with a `normal' SN Ia classification. (see Table \ref{tab:reclass} for the reclassification statistics and the second column of Table \ref{tab:DLR_cut_table}) for the updated numbers in each class (before any host contamination cuts).

We have also identified 59 SNe Ia that the initial DR2 typing did have a sub-classification for but we have overturned to be \uncIa. This gives a final total of 88 events with an \uncIa\ classification based on our re-classifications. The main causes of the lack of a sub-classification were either host contamination, where one or more critical features were not covered, or due to low S/N in the spectra. We discuss the impact of host galaxy contamination in further detail in Section \ref{sec:host_contamination}. We note that in this analysis, we are only concerned with their spectral classifications and do not consider any of their light curve properties such as $x_1$, $c$ or any galaxy properties in determining our classification and base our classifications purely on spectroscopic observations around peak.

\subsubsection{The normal to 91T-like spectral transition}
\label{sec:normal_to_91T}

\begin{figure}
\centering
\includegraphics[width=\columnwidth]{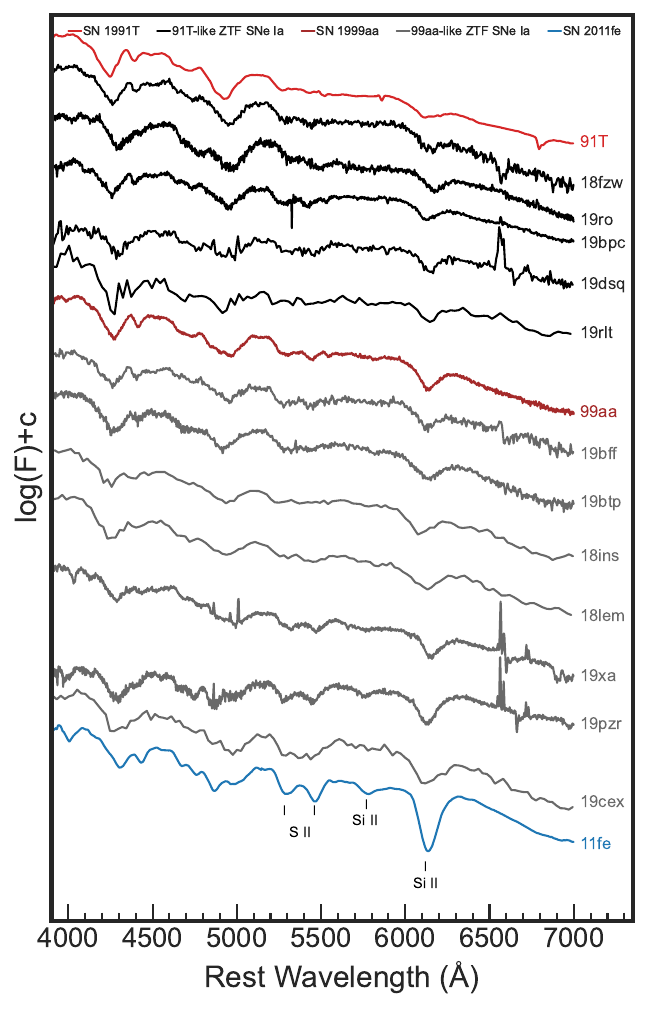}
\caption{Examples of `91T-like' (black) and `99aa-like' (grey) maximum light spectra are shown compared to the labelled spectra of SN 1991T, SN 1999aa and SN 2011fe at similar phases. Key features for identifying these SNe Ia are marked. Each spectrum has been corrected for MW extinction \citep{schlafly2011}. }
\label{fig:91t99aa99ee}
\end{figure}

In the normal to overluminous `91T-like' region, updated classifications were done by comparing the spectra in our sample at similar phases with the overluminous prototype SN 1991T, the normal SN 2011fe and a transitional event between these two classes, SN 1999aa (Fig. \ref{fig:91t99aa99ee}). `91T-like' SN Ia spectra around maximum light have weak \SiII\ features (6355 and 5972 \AA) compared to normal SN Ia spectra. \SII and \CaII H$\&$K features are considerably weaker than in a normal SN Ia, where \cite{phillips2022} showed the lack of \CaII H$\&$K feature itself is an identifier for true `91T-like' events. Objects with spectral properties between `91T-like' and `normal' SNe Ia are classified based on the transitional object SN 1999aa. While the \SiIIs\ feature of `99aa-like' SNe Ia near maximum light are stronger than `91T-like' SNe Ia, they are still on the lower side of `normal' SNe Ia. The classification of these transitional `99a-like' objects follow the closeness of pEW of \SiIIs\ to those in SN 1999aa. While some of these classifications are subjective, our classifications align closely with the method used to identify `91T-like' events from \citet{phillips2022} since near maximum light the pEW separation of \SiIIs\ itself is well established between these three sub-classes.

We visually inspected the 18 initially classified `91T-like' objects and reclassified 9 of them as bright transitional `99aa-like' SNe Ia based on the closer similarity of their \SiIIs\ feature strength to that of SN 1999aa than to SN 1991T or SN 2011fe. We confirm that in total seven of our sample are true `91T-like' SN Ia based on best-matching spectral comparisons to `91T-like' events. 39 SN are reclassified as `99aa-like', where 9 events where initially classified as `91T-like', 19 events as `normal' and 12 events as \uncIa\ (see Table \ref{tab:reclass}).

\subsubsection{The normal to 91bg-like spectral transition}
\label{sec:normal_to_91bg}

Similarly to the overluminous spectral transition, we have also visually inspected the sample to identify the diversity of subluminous events, spanning from `normal' to `91bg-like' SNe Ia. `91bg-like' events are generally recognised spectroscopically by the strong \TiII lines near 4300 \angstrom\ around peak brightness. A `W' shaped \SII feature is less clearly visible than in a `normal' SN Ia. \SiIIf\ and \SiIIs\ features and the \OI\ $\lambda$7773\ feature are especially prominent. Based on their spectroscopic similarities to those of SN 1991bg, 24 SNe in our sample are identified as `91bg-like'.

\begin{figure}
\centering
\includegraphics[width=\columnwidth]{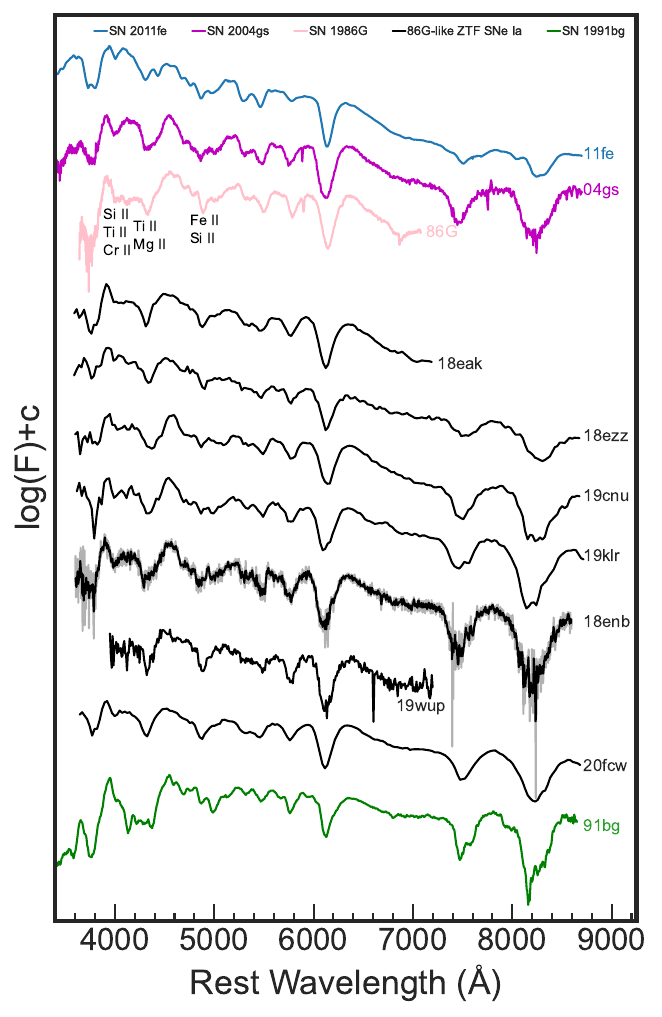}
\caption{A sub-sample of `86G-like' SN Ia spectra (shown in black) around maximum light are compared to those of SN 2004gs, SN 1986G, SN 1991bg and SN 2011fe at similar phases. Key features at identifying these SN Ia are shown. Each spectrum has been corrected for MW extinction. }
\label{fig:86G}
\end{figure}

SNe Ia identified as `86G-like' are transitional events between `normal' and `91bg-like' SN Ia \citep{phillips1987}. The spectra of SN 1986G had stronger \SiIIf\ and \SiIIs\ features than normal SNe Ia and some absorption due to \TiII\ in the 4300 \AA\ wavelength region but this was not as strong as seen in `91bg-like' events \citep{ashall2016}, which likely indicates an intermediate state between a `normal' and `91bg-like' SNe Ia. By following the spectroscopic characteristic differences mentioned above, we identified 11 SNe Ia in our sample as `86G-like' events, 3 of them were originally classified as `normal' and 8 as `91bg-like', demonstrating the difficulty in sub-classifying these events without careful comparison of relevant features. The strength of the \TiII\ feature is the main feature used to separate the `91bg-like' events and `86G-like' events. Examples of their maximum light spectra compared to other sub-classes are shown in Fig.~\ref{fig:86G}. 

\begin{figure}
\centering
\includegraphics[width=\columnwidth]{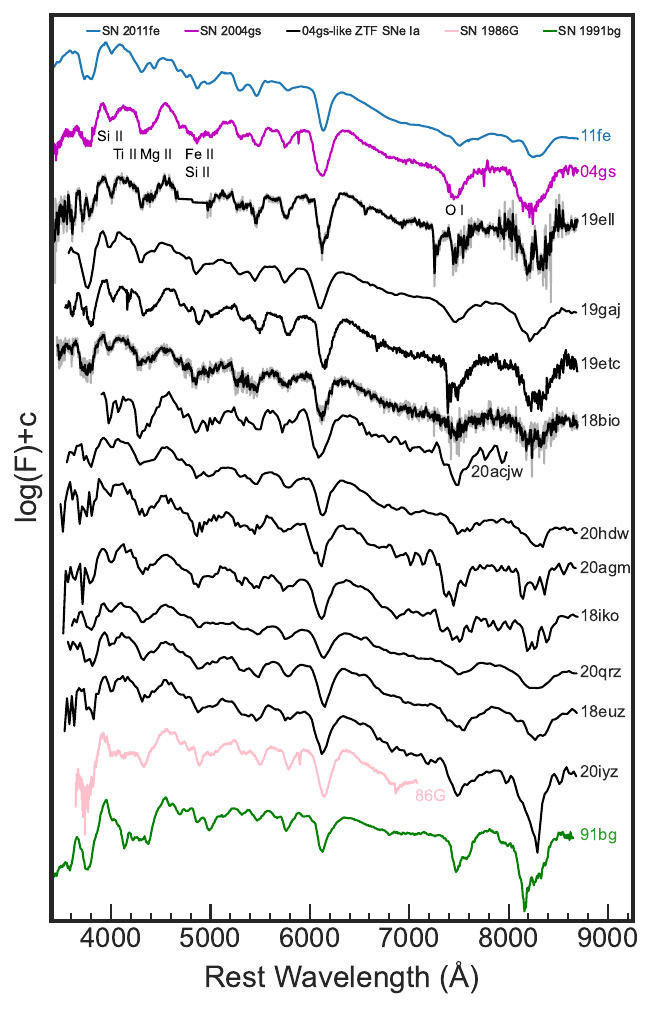}
\caption{A sub-sample of `04gs-like' (shown in black) ZTF DR2 SNe Ia around maximum light plotted with the spectra of SN 2004gs, SN 1986G, SN 1991bg and SN 2011fe at similar phases for comparison. Key features at identifying these SN Ia are shown in the figure. Each spectrum in the figure are corrected for MW extinction. }
\label{fig:04gs}
\end{figure}

In our sample, we have also identified SNe Ia that do not meet the spectroscopic criteria to be classified as `91bg-like' or `86G-like' but that are spectroscopically most similar to a faint normal SN Ia, SN 2004gs \citep{park2004, morell2004, blondin2012}. SN 2004gs was first discovered by \citet{park2004} and classified as a SN Ia by \citet{morell2004}. SN 2004gs is $\sim$ 10 kpc away from its S0 type host galaxy, PGC024286. SN 2004gs has been observed as a part of Center for Astrophysics Supernova Program (CfA) and its spectral and light curve parameters are published by \citet{blondin2012}, where it was classified as a `normal' SN Ia. However, SN 2004gs appears to be an intermediate state between a `normal' SN Ia and a transitional `86G-like' SN Ia. It shows significantly stronger \SiIIf, \OI\ $\lambda$7773 and \CaII NIR features and with weaker \TiII feature near 4300 \angstrom\ than seen in a `normal' SN Ia spectrum (see Fig.~\ref{fig:04gs}). The \TiII feature in the peak spectra is not as strong as a transitional (86G-like) SN Ia but still present, while the \MgII 4481 feature is more prominent in SN 2004gs. We identify 18 SNe in our sample as similar to `04gs-like' (some of which are shown in Fig.~\ref{fig:04gs}) based on the existence of stronger abundances of \SiIIf, \SiIIs, \OI\ $\lambda$7773, \CaII NIR than `normal' SNe with considerably weaker abundance of \TiII near 4300 \angstrom\ than `86G-like' SNe. They were all initially classified in the DR2 typing effort as normal SNe Ia. 

\begin{table}
\centering
\caption{Classifications of ZTF DR2 volume limited SN Ia sample with different \textit{d$_{DLR}$} cuts. The number of remaining SNe in each sub-type is shown.}
\label{tab:DLR_cut_table}
\begin{tabular}{lcccc}

\hline\\[-0.5em]
\textit{d$_{DLR}$} cut & no cut & $>$0.1 & $>$0.2 & $>$0.3 \\[0.15em]
    \hline\\[-0.8em]
    \hline\\[-0.5em]
Ia-unclear & 88 & 66 & \textbf{42} & 27\\[0.30em]
91T-like & 7 & 7 & \textbf{7} & 4\\[0.30em]
99aa-like & 39 & 37 & \textbf{34} & 27\\[0.30em]
normal & 295 & 280 & \textbf{257} & 236\\[0.30em]
04gs-like & 18 & 17 & \textbf{17} & 14\\[0.30em]
86G-like & 11 & 11 & \textbf{11} & 10\\[0.30em]
91bg-like & 24 & 24 & \textbf{24} & 22\\[0.30em]
    \hline\\[-0.8em]
    \hline\\[-0.5em]
Total & \multicolumn{1}{c}{482} & \multicolumn{1}{c}{442} & \multicolumn{1}{c}{\textbf{392}} & \multicolumn{1}{c}{340}\\[0.15em]
\hline\\[-0.5em]
\end{tabular}
\end{table}

\subsection{The impact of host galaxy contamination on SN Ia classification}
\label{sec:host_contamination}

\begin{figure*}
\centering
\begin{minipage}{0.33\textwidth}
  \centering
  \includegraphics[width=0.99\linewidth]{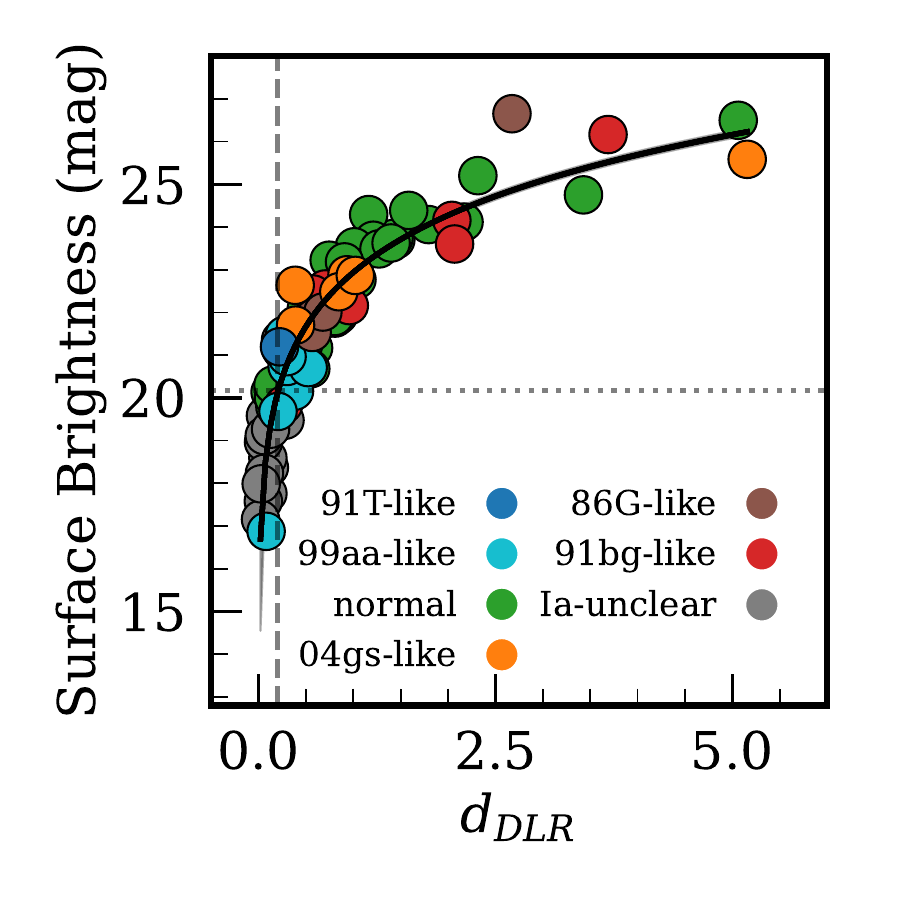}
\end{minipage}%
\hfill
\begin{minipage}{0.33\textwidth}
  \centering
  \includegraphics[width=0.99\linewidth]{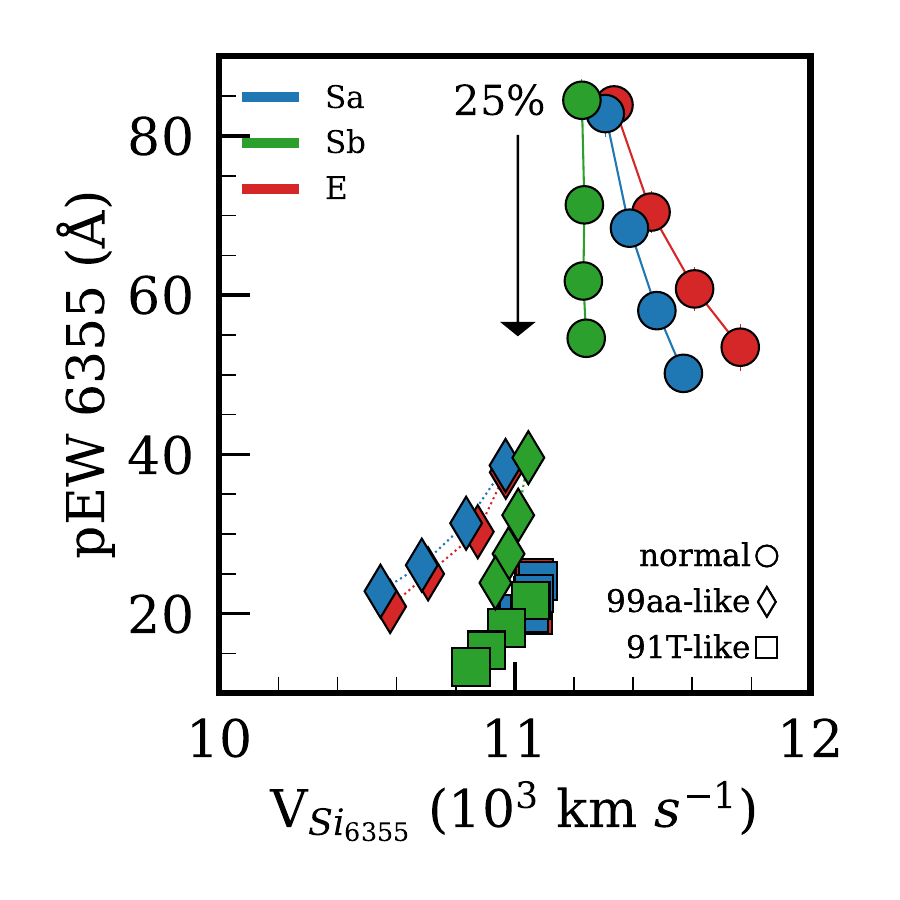}
\end{minipage}
\hfill
\begin{minipage}{0.33\textwidth}
  \centering
  \includegraphics[width=0.99\linewidth]{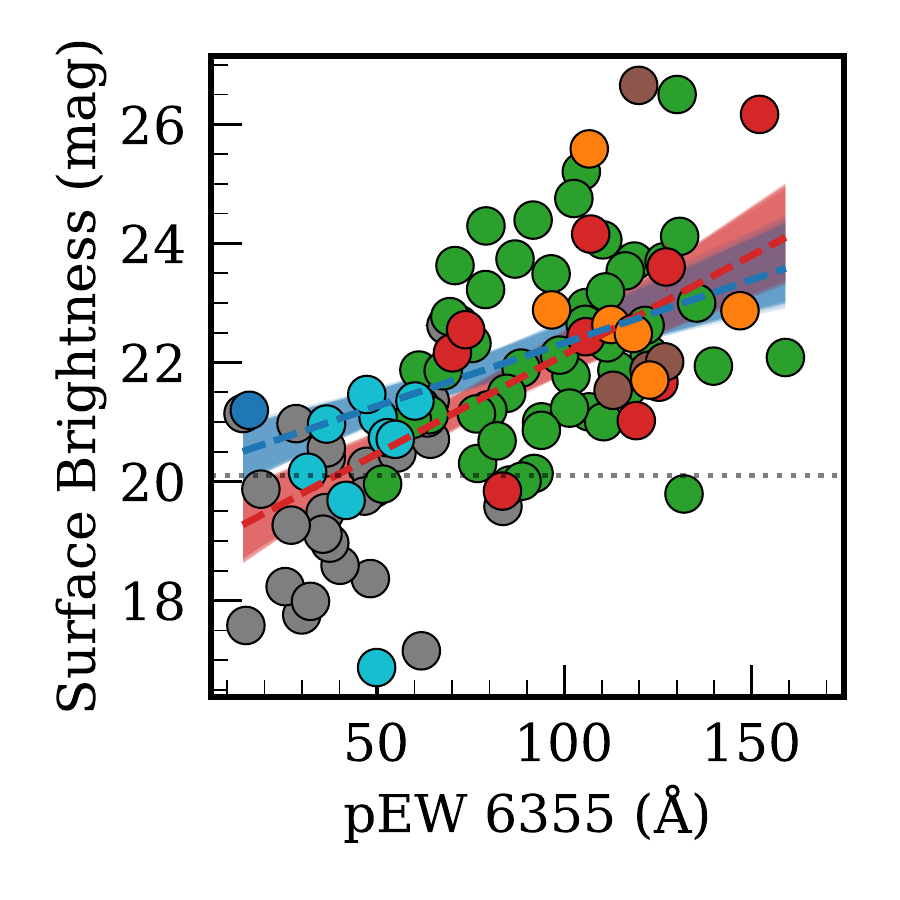}
\end{minipage}
\caption{\textbf{Left:} Local surface brightness compared with the \textit{d$_{DLR}$} of volume limited ZTF DR2 sample. The fit shown with a black solid line on the plot is used to estimate a \textit{d$_{DLR}$} value (vertical dashed line) based on the chosen surface brightness magnitude of 20.1 mag (vertical dashed line). \textbf{Middle:} We show the \SiIIs\ pEW against the velocity for three different subtypes of SNe Ia: 91T-like (represented by squares), 99aa-like (represented by diamonds), and normal (represented by circles). The data points for each SN subtype are shown with increasing levels of host galaxy light contamination (25, 50, 75, and 100 per cent) from top to bottom. Different colours represent different galaxy types used to simulate the contamination effects. This allows for the comparison of how various galaxy types influence the measured \SiIIs\ pEW for each SN subtype and how the pEW is affected by increasing galaxy contamination. \textbf{Right:} Local surface brightness plotted against the pEW of \SiIIs\ feature. The gray dashed line represents the equivalent value of \textit{d$_{DLR}$} where we applied our cut of \textit{d$_{DLR}$} = 0.2, based on the corresponding surface brightness magnitude of 20.1 mag, showed with the gray dotted line. The red and blue dashed line represents the regression line fitted to the all data and data with the \textit{d$_{DLR}$} cut, respectively. The red and blue shaded regions represent the confidence interval around corresponding regression lines. A strong correlation between the pEW of \SiIIs\ and local surface brightness and its effect on sub-typing is shown.}
\label{fig:foobar}
\end{figure*}

There are 88 SN Ia in our sample without a sub-classification, even after visual inspection. Some of these spectra are visibly contaminated by host galaxy light. Since 91T-like events have weak \SiIIs\ and \SiIIf, that could mean that it is preferentially classified as \uncIa\ if the S/N ratio is slightly lower or it is more contaminated by host galaxy light where it cannot be easily distinguished from a `normal' SN Ia. We have investigated a number of methods of quantifying the impact of host galaxy contamination on the SN Ia classifications and line measurements (pEW, velocity). 

Host galaxy properties of the ZTF DR2 sample is presented in \md{Smith et al.~(in prep.)}, where the software package \texttt{Host\_Phot}\footnote{\url{https://github.com/temuller/hostphot/tree/main}} was applied to Pan-STARRS DR2 \citep[PS1;][]{Chambers2016} images to obtain the Kron fluxes and the Directional Light Radius \citep[DLR;][]{sullivan2006,smith2012,gupta2016}. The DLR is the elliptical radius of a galaxy in the direction of the SN. The measure of the offset of a SN from the centre of its host galaxy that is scaled by the elliptical radius of the galaxy in the direction of the SN is called the \textit{d$_{DLR}$}. It is a measure of the SN to host galaxy centre separation corrected for the differing sizes of galaxies. The \textit{d$_{DLR}$} distribution of the ZTF DR2 sample ranges from zero to seven but peaks at a value of $\sim$0.2, with most SNe Ia having \textit{d$_{DLR}$} values of less than three.

The local surface brightness of the galaxy at the position of the SN can also be estimated as a measure of the likely host contamination in our spectral measurements. These values were estimated for SNe Ia in the ZTF DR2 using a galaxy image decomposition involving four morphological models, bulge, disk, bulge+disk and bulge+bar+disk \md{(Senzel et al.~subm.)}. The model fits were performed in surface brightness space (mag/arcsec$^2$) and the local surface brightness values used in this paper are sampled from the models at the location of the SN. The galaxy models undergo several quality control steps during the fitting procedure, ensuring that only well-fit and physically accurate models make it into the final sample.  From the original 3628 SNe Ia in ZTF DR2, the final sample consists of 724 modelled galaxies. After the applied selection criteria (see Sec.~\ref{sec:sampleselect}), 115 out of 482 SN Ia have surface brightness values ($\sim$24 per cent).

The local surface brightness is likely a better measure than the \textit{d$_{DLR}$} of the contamination of the SN spectra but unfortunately is only available for 24 per cent of the SNe Ia in our final sample. We have compared our \textit{d$_{DLR}$} and local surface brightness values in the \textit{g} band for the SNe Ia that have both measurements (see Fig.~\ref{fig:foobar}, left). We identify a clear trend between both quantities, where the larger the \textit{d$_{DLR}$} (larger corrected offset from the host centre), the fainter the underlying local surface brightness. A power law model for \textit{d$_{DLR}$} and local surface brightness relation was found to fit the quantities well. 

We have colour coded in the SNe Ia in Fig.~\ref{fig:foobar} and find that the \uncIa\ sub-class is much more highly represented at small \textit{d$_{DLR}$}/bright local surface brightness compared to normal events or other sub-classes. This suggests that the inability to sub-classify is likely driven significantly by the contamination from host galaxy light. However, since some sub-classes may have host galaxy dependencies, e.g., a general preference for 91T-like events to occur in star-forming host galaxies \citep{phillips2022}, we want to bias our sample as little as possible. Therefore, we would like to apply a balanced cut on the \textit{d$_{DLR}$}/surface brightness values that removes a large percentage of the \uncIa\ that occur in the inner brighter regions of their host galaxies but does not remove too many SNe Ia from our sub-classes with smaller numbers of events. 

To do this, we test the impact of removing events with \textit{d$_{DLR}$} less than 0.1, 0.2, and 0.3 compared to no cut on the sample on the different sub-types (Table \ref{tab:DLR_cut_table}). By placing a cut to remove events that have \textit{d$_{DLR}$}$<$0.1, we would remove $\sim$25 per cent of the \uncIa\ sample, only $\sim$5 per cent of the `normal' SN Ia sample, two `99aa-like' event and one `04gs-like' event, which is equivalent to 2 per cent of the other sub-types. Increasing this cut to \textit{d$_{DLR}$}$<$0.2 would remove $\sim$52 per cent of the \uncIa\ classification, $\sim$12 per cent of the `normal' SN Ia sample, and $\sim$6 per cent from the rarer sub-classes. Increasing the cut to remove objects with \textit{d$_{DLR}$}$<$0.3 removes $\sim$70 per cent of the \uncIa, $\sim$20 per cent of the `normal' SN Ia, and $\sim$20 per cent from the rarer sub-classes. At all cut levels, a much more significant percentage of events without a sub-classification (\uncIa) is removed compared to the other sub-classes. For our final sample, the choice of \textit{d$_{DLR}$} is somewhat arbitrary but we choose to exclude all the events with \textit{d$_{DLR}$}$<$0.2 based on the minimal impact it has on the rarer sub-types while removing nearly half of the \uncIa. Some of the remaining \uncIa\ in the sample are likely contaminated by host galaxy features but may also not have the wavelength coverage for identification of features necessary for sub-classification or low SNR at the wavelengths of key features. This \textit{d$_{DLR}$} cut of 0.2 is equivalent to a local surface brightness of 20.1 mag based on our power law model fit to the relation (left panel of Fig.~\ref{fig:foobar}). It is this final sample of 392 SNe Ia (Table \ref{tab:DLR_cut_table}) that is considered in the subsequent analysis.

\subsection{The impact of host galaxy contamination on \SiII\ feature measurements}
\label{sec:host_contamination_SiII}

We demonstrated the direct impact of host contribution on SN Ia spectral sub-typing in Section \ref{sec:host_contamination} and removed those events that are most heavily contaminated. However, it is likely that host contamination is still present in SNe Ia in our sample and we wish to determine what impact this may have on the measured spectral parameters (velocity, pEW). We tested this by artificially injecting host galaxy contamination into test SN Ia spectra from our sample. We did this by taking spectra that were visually uncontaminated by host light of a normal, a `91T-like', and a `99aa-like' SN Ia and combining with template galaxy spectra of two spirals (Sa, Sb) and an elliptical (E) from \cite{kinney1996}. Each SN spectrum was combined with each galaxy template at varying flux ratios of galaxy contamination of 0.25, 0.5, 0.75 and 1 and then the combined spectra were fit with our spectral line analysis code to determine the parameters of the \SiIIs\ features. 

In middle panel of Fig.~\ref{fig:foobar}, we show the pEW of the \SiIIs\ feature against its velocity for the three SN Ia sub-classes with varying host contamination of the three different galaxy templates. We observed a considerable decrease in the measured pEW of each sub-class as the host galaxy contamination was increased, ranging from a 17 per cent change to almost 50 per cent in the normal SN Ia for the most contaminated spectrum. The velocity measurement changes were smaller with an increase of up to 450 \kms\ for the normal SN Ia and decrease in the 99aa-like event of 500 \kms. A decrease in pEW measurements in SN Ia spectra was also determined by \cite{nordin2011} due to host galaxy contamination. 

In the right panel of Fig.~\ref{fig:foobar}, we show the \textit{g}-band local surface brightness against \SiIIs\ pEW for SNe Ia in our sample \textit{before} the \textit{d$_{DLR}$} cut of 0.2. A linear regression fit to the data is shown with red dashed line on the right panel of Fig.~\ref{fig:foobar}. The events with an \uncIa\ classification cluster below a pEW of $\sim$70 \AA. Based on our \textit{d$_{DLR}$} cut that removes values below 0.2, this is equivalent to a value of 20.1 mag for the local surface brightness. We then remove all SNe Ia with brighter local values than 20.1 mag, where the blue dashed line on the right panel of Fig.~\ref{fig:foobar} shows the linear regression fits for before and after the removal of the objects. After removing objects brighter than 20.1 mag, the significance level between the surface brightness and the pEW of \SiIIs\ seemed to decrease by approximately two sigma. This suggests that the events heavily contaminated by host galaxy light are driving this correlation. However, our tests of spectral contamination indicate that the rest of the sample may also be affected, leading to smaller measured values of the \SiIIs\ pEW than those that are intrinsically present.

\section{Results and Discussion}
\label{sec:results}

In Section  \ref{sec:sub-classes}, we present the `Branch' classification diagram \citep[][]{branch2006} of the ZTF SN Ia DR2 sample based on their pEW values. We also discuss the relative rates of SNe Ia in each of the subclasses and their key differences. In Section \ref{sec:wang-pEW-Vel} we present the pEW and velocity distributions of `normal' SNe Ia from the `Wang' classification scheme (based on pEW and velocity of the \SiIIs\ feature) and investigate the properties of NV and HV SNe Ia. The spectral properties and \textsc{tardis} analysis of faint `normal' SNe Ia are presented in Section \ref{sec:04gs_cont}. In Section \ref{sec:LCprop}, the connection between the spectral properties of SNe Ia and their SALT2  light curve properties are presented.

\begin{figure*}
\centering
\includegraphics[width=17cm]{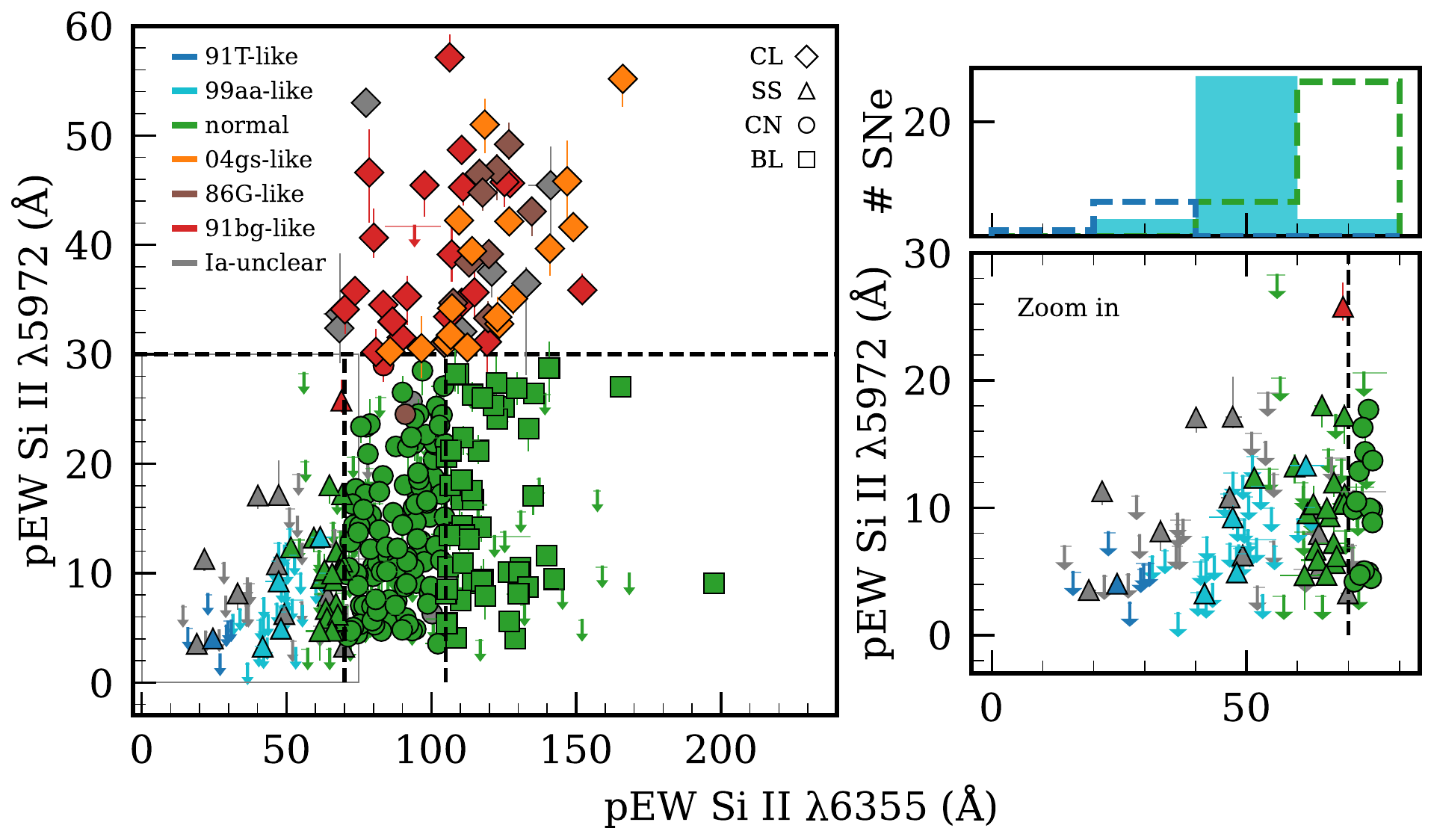}
\caption{\textbf{Left:} The pEW of \SiIIf\ plotted against the pEW of \SiIIs\ based on sub-types of SNe Ia taken from \cite{branch2006} for the 392 SNe Ia from the ZTF DR2 sample with \textit{d$_{DLR}$} greater than 0.2. Branch sub-type zones are shown with the black dashed lines following the limits from \citet{folatelli2013}, with `SS' for shallow Silicon, `CN' for core normal , `BL' for broad line and `CL' for cool events. Filled circles with different colours represent the SN Ia types. Upper limits shown on the plots are plotted as downward arrows (see Section ~\ref{sec:sec2velocities}). \textbf{Right:} Zoom-in plot of the SS SN Ia region along with the histogram on top of the number of events of the `91T-like', `99aa-like' and `normal' SN Ia sub-types in this region, showing the transition from `91T-like' (dark blue) to `99aa-like' (cyan) to `normal' (green) with increasing pEW of \SiIIs.}
\label{fig:Branch}
\end{figure*}

\subsection{Branch Classification}
\label{sec:sub-classes}

In Fig.~\ref{fig:Branch}, we present the pEW of the \SiIIf\ feature against the pEW of the \SiIIs\ feature for spectra in the phase range of -5 to +5 days with respect to the time of maximum light, $t_0$. SNe Ia with a \textit{d$_{DLR}$} less than 0.2 have been removed. In cases where multiple spectra are available for a given SN within the phase range, the one nearest to maximum light is chosen.

It is well established that the pEWs of the \SiIIf\ and \SiIIs\ features are strongly correlated in the identification of SN Ia sub-types \citep[][]{branch2006}, where they can be separated into four spectral sub-classes: `core normal' (CN), `shallow silicon' (SS), `broad line' (BL), and `cool' (CL), based on the strength ratio of these features. We follow the same criteria as \cite{folatelli2013} to assign the `Branch' sub-classes. A clustering analysis of the Branch diagram parameters using a Gaussian mixture model suggested that these four broad classes are well-defined statistically \citep{burrow2020}.

In our sample, we have 35 SS, 122 CN, 51 BL and 56 CL coming from direct measurements of the features. Using our three-sigma limits on the pEW of \SiII\ 5972 \AA\ feature, we can assign a additional objects to the subclasses of 69 SS, 39 CN, 19 BL, and one CL. Out of 392 SNe Ia, $\sim$26$\%$ have a SS sub-class, $\sim$41$\%$ have a CN sub-class, $\sim$18$\%$ have a BL sub-class and $\sim$15$\%$ have a CL sub-class. \citet{blondin2012} analysed SN Ia spectra obtained by the Centre for Astrophysics (CfA) Supernova Program and assigned Branch classifications using spectra around maximum light. Out of 218 SNe Ia in their sample $\sim$15$\%$ have a SS sub-class, $\sim$40$\%$ have a CN sub-class, $\sim$28$\%$ have a BL sub-class and $\sim$17$\%$ have a CL sub-class. In another work, \citet{folatelli2013} presented the optical spectra of SNe Ia with the spectra obtained by the Carnegie Supernova Project (CSP; \citet{hamuy2006}). Out of 78 SNe Ia at maximum light in their sample, $\sim$13$\%$ have a SS sub-class, $\sim$32$\%$ have a CN sub-class, $\sim$28$\%$ have a BL sub-class and $\sim$27$\%$ have a CL sub-class. Assuming Poissonian uncertainties, on the relative percentages, our relative rates are broadly comparable to those of CfA and CSP. However, our SS rate is slightly higher, which may be due to the CfA and CSP studies not including events with only upper limits on their \SiII\ 5972 \AA\ feature but their spectra are generally higher S/N than the mean of our sample.

In Fig.~\ref{fig:Branch}, we also show the spectral types (`91T-like', `99aa-like', `normal', `04gs-like', `86G-like', `91bg-like', and `Ia-unclear') we have identified for the 392 SNe Ia in Section \ref{sec:template_matching}. The `normal' SNe Ia fall generally fall into the CN and BL region, with some at the boundary of the \SiIIs\ pEW between SS and CN (discussed more in Section 
\ref{sec:91t_rare}). The `91bg-like', `86G-like', and `04gs-like' SNe Ia are predominantly found in the CL region of Fig.~\ref{fig:Branch}, while the `91T-like' and `99aa-like' events are found in the SS region. These clusters of spectroscopic sub-classes in the Branch diagram are not unexpected since the \SiII\ features are some of the main features used to sub-classify SN Ia spectra. 

\subsubsection{How rare are 91T-like events?}
\label{sec:91t_rare}

The SS Branch sub-class (those with both weak \SiIIf\ and \SiIIs\ features) is made up of $\sim$33 per cent `99aa-like', $\sim$32 per cent `normal' SNe Ia, $\sim$28 per cent \uncIa, and $\sim$7 per cent `91T-like'. All `91T-like' and `99aa-like' SNe Ia in our sample are in the SS region. While we see similar rates of `normal' and `99aa-like' SNe Ia in the SS region (see right panels of Fig.~\ref{fig:Branch}), true `91T-like' SNe Ia remain rare at $\sim$7 per cent. All `91T-like' SNe Ia have shallower \SiIIs\ features than the `99aa-like' events. The `91T-like' SNe Ia all have \SiIIs\ pEW values below 31 \AA, while the `99aa-like' SNe Ia have values in the range of 32 to 63 \AA. Only four `99a-like' SNe Ia have a pEW \SiIIs\ value higher than 55 \AA\ (55, 60, 62 and 63 \AA). Normal SNe Ia start to dominate the population with \SiIIs\ pEW above $\sim$55 \AA. Conversely, only two `normal' SNe Ia have pEW \SiIIs\ values lower than $\sim$55 \AA, at 51.5$\pm$$^{\text{0.4}}_{\text{0.6}}$\ and 54.5$\pm$$^{\text{1.8}}_{\text{2.8}}$\ \AA.

Fig.~\ref{fig:Branch} shows that most SNe Ia without a sub-classification (\uncIa) are on the lower end of the pEW \SiIIs\ distribution. Approximately 70 per cent of the 42 SNe Ia without a sub-classification are in the SS region. We have shown in Section \ref{sec:host_contamination} that significant host contamination has the impact of decreasing the measured pEW of the \SiII\ lines, while also making sub-classification more difficult in general. Therefore, it is unsurprising that the majority of events without a sub-classification are among the events with the weakest \SiII\ pEW values. In the most extreme case where the host galaxy contamination is comparable to the SN flux, the pEW of the \SiIIs\ feature may be up to 35 \AA\ weaker due to host contamination than the intrinsic SN measurement. This could result in many of the unclassified \uncIa\ class moving into the core normal region on the Branch diagram (see bottom right panel of Fig.\ref{fig:Branch}).

The rareness of the `91T-like' events can not be explained by the host contamination because significant host contamination decreases the pEW. Even if we assume all \uncIa\ that have similar \SiIIs\ pEW values to those of the `91T-like' SNe ($<$31 \AA) are also `91T-like', the rate is still less than 5 per cent of all SNe Ia in our sample. 

\cite{phillips2022} identified a sample of `91T-like' events based on their \SiIIs\ pEW as a function of phase, the time of maximum in the \textit{i} band relative to the \textit{B} band, and the presence of a clear secondary maximum in the \textit{i} band. Their \SiIIs\ pEW cutoff was $<40$ \AA\ at \textit{B}-band maximum to $<50$ \AA\ at +10 days relative to maximum light. Our maximum-light ($-$1 to +5 d) `91T-like' SNe are defined by the visual inspection method discussed in Section \ref{sec:normal_to_91T} and broadly in agreement with \cite{phillips2022}. They identified that their `91T-like' events are overluminous in their Hubble residuals compared to other SS and CN SNe with similar light curve decline rates. 

Several `normal' SS SNe Ia are spectroscopically quite similar to the SNe Ia in lower limits of CN SNe Ia, showing a continuity in the pEW of \SiIIs\ feature from SS to CN class. Hence, the defined separation at a pEW of \SiIIs\ at 70 \AA\ between SS and CN becomes less relevant than the separation between `99aa-like' and `normal' SN Ia at $\sim$ 55 \AA. A recent study from \citet{chakraborty2023} shows that `99aa-like' SN Ia can be separated into two groups, one showing similar LC properties to `91T-like' SN Ia and other closer to brighter end of `normal' SN Ia, suggesting the later one bridging a continuity from normal to bright end of SN Ia. \citet{obrien2023} did \textsc{tardis} based analysis on `91T-like' SN Ia and `normal' SN Ia. Through the effect of abundances and ionisation states of intermediate-mass elements in the spectra of these sub-types, it is concluded that continuity is seen between the `normal' and `91T-like' SN Ia.

\subsubsection{The diversity of `cool' events}
The CL SN Ia sub-class of \cite{branch2006} consists of SNe Ia with deeper \SiIIf\ absorption (Fig.~\ref{fig:Branch}), often associated with low-luminosity events. The `91bg-like' and `86G-like' events are seen to dominate the CL class, but there are faint normal SNe Ia (`04gs-like') that fall into this region in Fig.~\ref{fig:Branch} as well. We have visually inspected these events and find that all 17 `04gs-like' SNe Ia in the CL class have the strong \SiIIf\ of `91bg-like' and `86G-like' events but have significantly less \TiII $\lambda$4300. These faint SN Ia also have stronger \CaII NIR and \OI\ in their spectra than `normal' SNe Ia, populating the region between `normal' and the `86G-like` events. While these faint SNe Ia do not show any preference in Fig.~\ref{fig:Branch}, other than all being in CL class, they do exhibit a clear difference in their light curve properties (see Section \ref{sec:LCprop}). Seven SNe Ia in the CL region did not have key features in their spectra or were still host contaminated, making us unable to sub-classify them.

\subsubsection{A continuum from `core normal' and to `broad-line'}
\label{sec:CN_and_BL}

In Fig.~\ref{fig:Branch}, the Branch classification of CN is dominated by SNe Ia that are also spectroscopically classified as `normal' SNe Ia ($\sim$95 per cent). This suggests that the vast majority of `normal' SNe Ia can be identified by solely the pEW of their \SiIIs\ and \SiIIf\ features. There is only one `91bg-like' event and one `86G-like' event falling in this CN group. Both of these events displayed significant host contamination in their spectra (although they have \textit{d$_{DLR}$} values above our cut-off of 0.2), diminishing the \SiIIf\ feature, hence lowering the pEWs of these SNe Ia (see Section \ref{sec:host_contamination}) from likely CL to CN. 

The BL SN Ia sub-class is exclusively SNe Ia that are visually spectroscopically classified as `normal' SNe Ia. This is likely because their most common spectral match in the \textsc{snid} database is the BL SN Ia, SN 2002bo \citep{benetti2004_02bo,stehle2005_02bo}. The most extreme BL in our sample is SN 2018ccl, with a very high \SiIIs\ pEW  of $\sim$197 \AA. This extreme case is similar to SN 1984A \citep{branch2006}, which had a \SiIIs\ pEW  of $\sim$200 \AA. 

Previous studies have found that a significant fraction of SNe Ia possess secondary, high-velocity absorption components, predominantly in the \SiIIs\ and \CaII NIR features \citep{hatano1999,quimby2006,childress2013,maguire2014} up to maximum light. The findings of \md{Harvey et al.~(subm.)} suggest that up to 42$\pm$12 per cent of BL SN Ia classifications are due to the presence of an additional high-velocity component that broadens the line. \citet{burrow2020} used Gaussian mixture models to investigate the Branch diagram from their data and showed that when the \SiIIs\ velocity is taken into account in the analysis, the BL group is nearly distinct from the other Branch sub-classes, which further supports the case of the BL class being highly affected by the HV components since the combined velocity would be higher when two blended components are present.

\subsection{`Wang' Classification and Velocity-pEW Relation}
\label{sec:wang-pEW-Vel}

\begin{figure}
\centering
\includegraphics[width=\columnwidth]{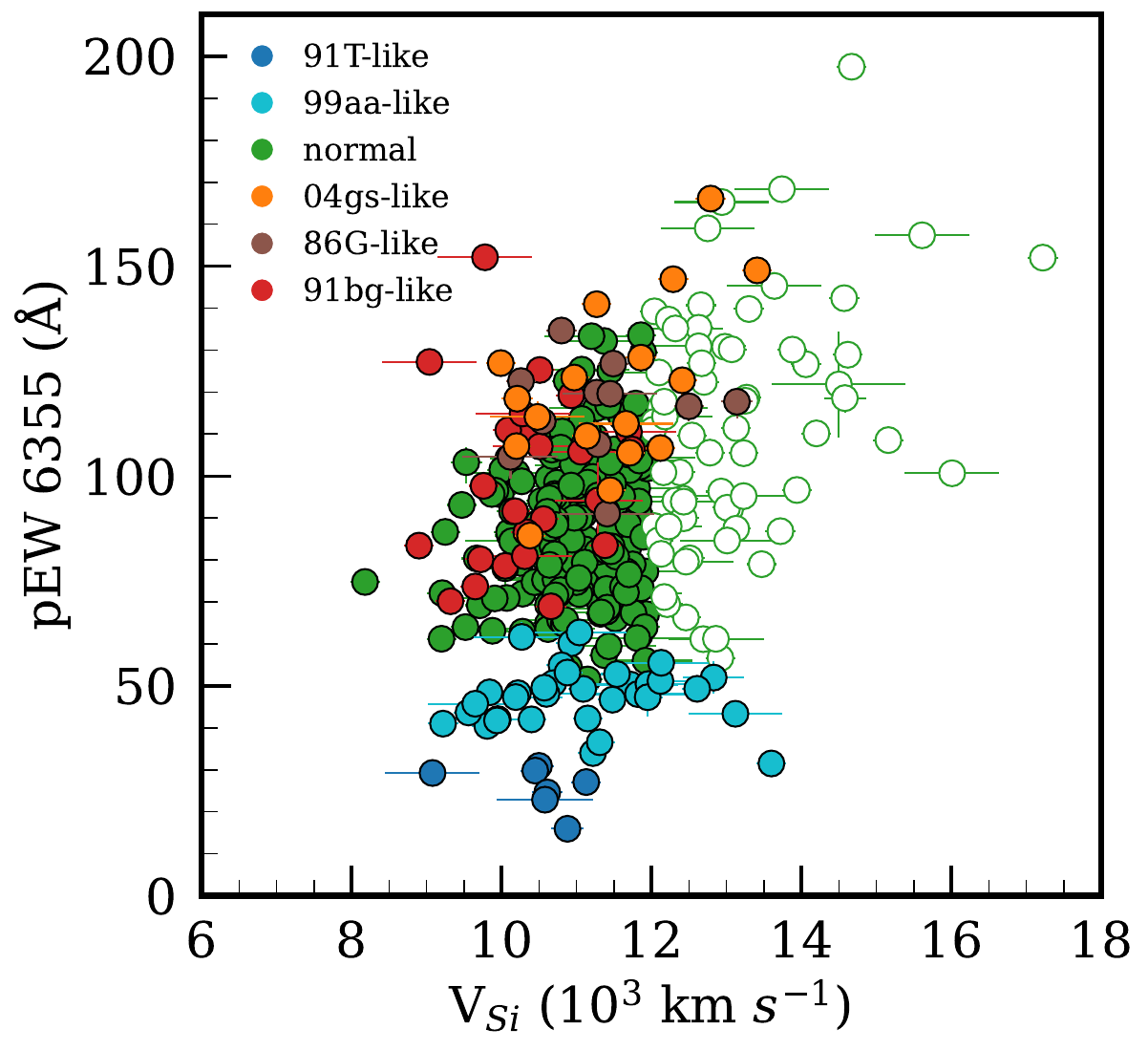}
\caption{Pseudo equivalent widths of the \SiIIs\ feature plotted against the velocity of the same line ($v_{\mathrm{Si}}$) near maximum light ($-$5 to $5$ days). Where several measurements exist for a given SN in the phase range used, we selected the one closest to maximum light. Different colours represent different sub-types used in this study. Green filled and open circles represent the NV SNe Ia and HV SNe Ia within the `normal' SNe Ia in our sample, respectively. }
\label{fig:vel_pew}
\end{figure}

The `Wang' classification scheme introduced by \cite{wang2009_may_a} of normal velocity (`NV', $v_{\mathrm{Si}} \lesssim$ ~12,000 km s$^{-1}$) and high velocity (`HV', $v_{\mathrm{Si}} \gtrsim$ ~12,000 km s$^{-1}$) is solely based on the velocity of the \SiIIs\ absorption feature around the peak of the SN light curve (Fig.~\ref{fig:vel_pew}). This classification only applies to normal SNe Ia, and all other sub-types, including the \uncIa, are excluded from the classification scheme. 257 out of 392 SN Ia in our sub-sample are classified as `normal', and we assigned 191 SNe Ia as `NV SN Ia' (74$\%$) and 66 SNe Ia as `HV SNe Ia' (26$\%$). The mean velocity and the mean pEW 6355 of the `NV SNe Ia' in our sample are $-$ 10,954$\pm$47 \kms\ and 86$\pm$1 \angstrom, respectively (Fig.~\ref{fig:vel_pew}). The mean velocity and the mean pEW 6355 of the `HV SNe Ia' in our sample are $-$ 13,072$\pm$130 \kms\ and 103$\pm$4 \angstrom, respectively. These are broadly consistent with the results of \cite{blondin2012}, where the mean velocity and the mean pEW 6355 of the NV SNe Ia were estimated as $\sim-$11,000 \kms\ and $\sim$95 \angstrom. Small differences are not unexpected given differences in the origins of host galaxy redshifts (\textsc{snid} compared to spectroscopic), the impact of smoothing, as well as our wider spectral phase range of $\pm$5 d from maximum light, compared to $\pm$3 d from maximum light  in \cite{blondin2012}. Since in our sample we have more early spectra (Fig.~\ref{fig:hist}) and for `normal' SNe Ia the earlier spectra tends to have smaller pEW values \citep{folatelli2013} than the after maximum light ones, this could explain the average pEW of \SiIIs\ in our sample being smaller by a few angstroms compared to \cite{blondin2012}.

As can be seen in Fig.~\ref{fig:vel_pew} `99aa-like' objects have a similar velocity distribution to `normal' SNe Ia, with an average \SiIIs\ velocity of 10,935$\pm$184 \kms\ but a lower average pEW of 46$\pm$1 \angstrom. `91T-like' objects in our sample on average slightly slower than the `normal' and `99aa-like' objects with an average \SiIIs\ velocity of 10,676$\pm$247 \kms\ and have smaller pEWs of 26$\pm$2 \angstrom. `91bg-like' objects also show similar velocities with an average \SiIIs\ velocity of 10,266$\pm$156 \kms\ and higher pEWs of 91$\pm$4 \angstrom.

\begin{figure}
  \begin{minipage}{\linewidth}
    \centering
    \includegraphics[width=\linewidth]{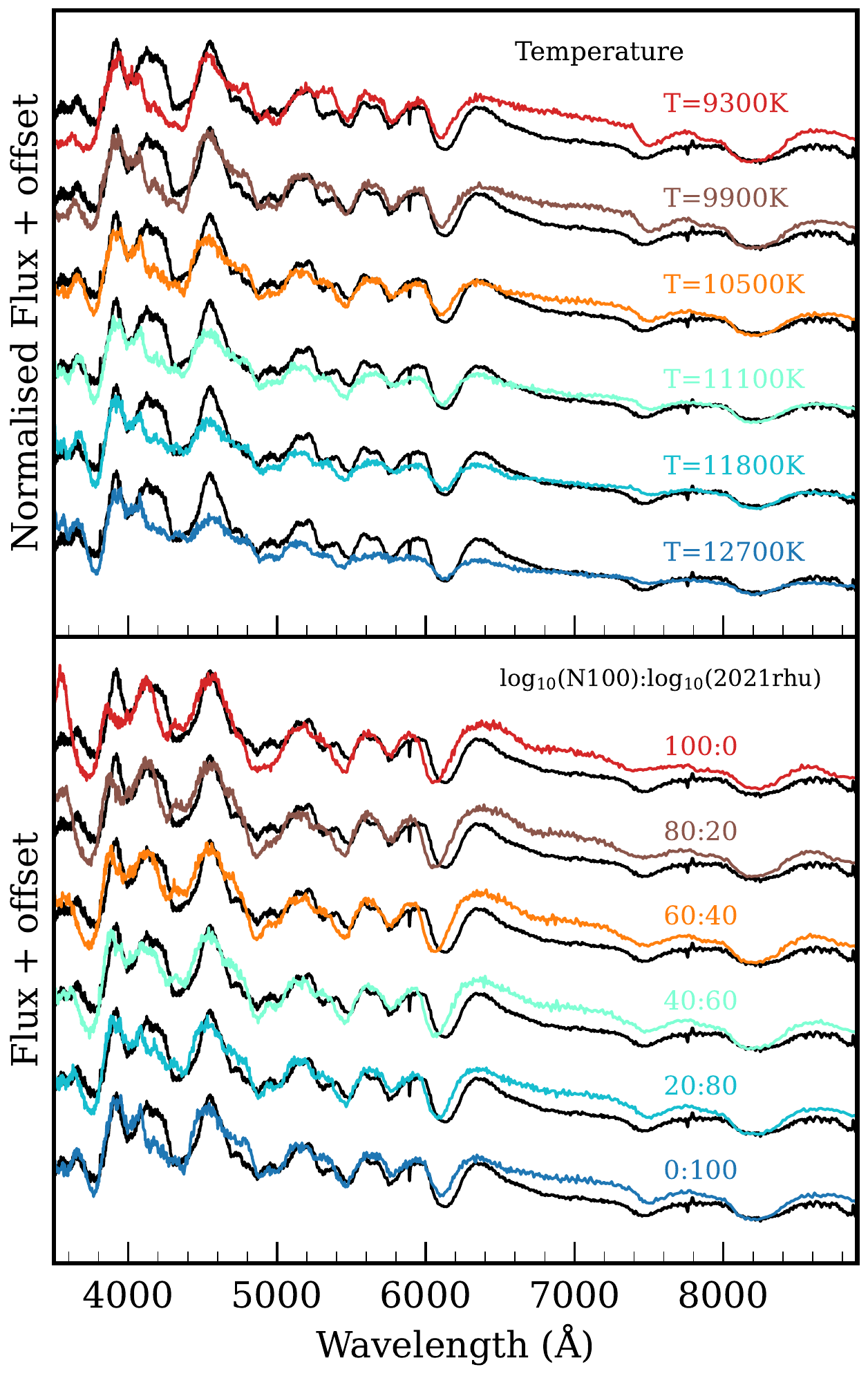}
  \end{minipage}
  \caption{\textbf{Top:} Custom model spectra created for SN 2021rhu compared to a +0.3~d relative to \textit{g}-band peak spectrum of SN 2004gs (black solid line). Each colour represent a different temperature setting while keeping the rest of the initial conditions as same. Spectra shown are normalised by their value at 4000 \angstrom\ to better line up the \TiII and \SiII\ features. It is seen that the varying temperature impacts the strength of the \SiIIf\ features but has no considerable impact on the \TiII feature. \textbf{Bottom:} A set of spectral configuration of N100 (red) model and the custom model created for SN 2021rhu (dark blue model). The ratios of each model mixed is shown next to each model spectrum (calculated in log space). The best match of SN 2004gs is with the 40:60 model, which implies that the abundances of the 04gs-like faint SNe Ia are in between the transitional 86G-like and the normal SNe Ia. SN 2004gs is less luminous than SN 2021rhu, slightly cooler and therefore, has a stronger \SiIIf\ feature.}
    \label{fig:tardis}
\end{figure}

\subsection{Are 04gs-like events a continuum from normal SNe Ia?}
\label{sec:04gs_cont}


To determine if the intermediate spectral features of SN~2004gs between those of normal and sub-luminous (`86G-like') SNe Ia can be achieved by temperature or abundances changes, we use the 1D Monte Carlo radiative transfer code \textsc{tardis} \citep{Kerzendorf2014, Kerzendorf2020}. The input parameters for \textsc{tardis} consist of the luminosity, the time elapsed since the explosion, the photospheric velocity, and profiles of abundance and density. \textsc{tardis} works under the assumption that a distinct, well-defined photosphere emits a blackbody continuum, bridging the transition between the optically thick and thin regions of the ejected material.\textsc{tardis} is commonly used and has proven to be successful at applying constraints to the broad range of SN ejecta coming from different sub-types such as `91T-like' SNe Ia \citep{obrien2023}, normal SNe Ia \citep{obrien2021, ogawa2023}, `02cx-like' SNe Ia \citep{magee2016, srivastav2020}, and `86G-like' SNe Ia \citep{harvey2023}. By varying the input parameters, namely the luminosity and abundance profile, we aim to have a self-consistent model that can produce the primary key features (\TiII, \SiII, \OI, and \CaII NIR) of a photospheric-phase spectrum (+0.3~d relative to \textit{g}-band peak) for SN~2004gs.

Firstly, we investigate the effect of changing the photospheric temperature, choosing our abundance and density profiles to be the custom SN~2021rhu model from \citet{harvey2023}. This model was chosen as it reproduces well  the \TiII evolution in the 86G-like event, SN~2021rhu, which appears to be a stronger variation of the peculiarity seen in SN~2004gs. With the photospheric velocity at 10,500~km s$^{-1}$ and the time since explosion at 16.8 d (input parameters for the $-$1.4 d spectrum of SN~2021rhu), we vary the photospheric temperature through the requested luminosity parameter to evaluate the effect upon the morphology of the \TiII feature at $\sim$4300 \AA. We run these simulations at six temperatures in the range of 9300 to 12700 K (Fig.~\ref{fig:tardis}, top panel). While there is a large evolution with the changing temperature in the \SiIIf\ feature and therefore, the \SiII\ 5972 to \SiII\ 6355 \AA\ ratio, there is little evolution in the \TiII feature, which fails to match the faint \TiII seen in SN 2004gs spectra, while having a strong \SiIIf\ feature.

We subsequently investigate various blended models between this custom model for SN~2021rhu and the delayed detonation N100 model \citep{ropke2012, seitenzahl2013}. The time since explosion is once again taken as 16.8 d, and the photospheric velocity as 10,500~km~s$^{-1}$, with the luminosity parameter fixed as log$_{10}$\textit{(L/L$_\odot$) =} 9.2  to align with the spectrum of 2004gs; lower than the log$_{10}$\textit{(L/L$_\odot$) =} 9.3 for SN~2021rhu. We start with the pure N100 model, mixing in more of the SN~2021rhu abundance profile with each step but leaving the temperature fixed. In each step we mix in 20 per cent more of the SN~2021rhu abundance profile in log space until it becomes the pure SN~2021rhu model. The resulting abundance profile, \textit{X}, can be expressed as,
\begin{equation}
   \log_{10}(X) = A \log_{10}{X_{\text{N100}}} + (1-A) \log_{10}{X_{\text{2021rhu}}}
\end{equation}
where \textit{A} is the fraction of the N100 abundance profile, $X_{\text{N100}}$ is the abundance profile of the N100 model, and $X_{\text{2021rhu}}$ is the abundance profile of the SN 2021rhu model.

As the temperature remains fixed between these simulations (only varying the abundances between the pure N100 and the SN 2021rhu-matched models) the \SiII\ ratio stays relatively constant for each mixing step, matching the observed spectrum closely. However, due to the changing abundances, there is significant variation in the \TiII feature, with the best match for the  weaker \TiII line seen in SN~2004gs compared to SN~2021rhu coming from the 40:60 model (Fig.~\ref{fig:tardis}, bottom panel). This model consists of 40 per cent of the N100 model abundances and 60 per cent of the SN~2021rhu model bundances (scaled in log space), and produces the strong \SiIIf\ feature and relatively strong \OI\ and \CaII NIR features, along with the relatively weaker \TiII 4300 \AA\ feature to match the SN~2004gs spectrum well. Our findings suggest that variations in temperature alone are insufficient to produce the spectral properties seen in faint normal SNe Ia compared to normal SNe Ia. A combination of both temperature and abundance profile is key in the formation of these events and suggests that there is a potential continuum from normal to faint SNe Ia. 

\cite{nugent1995} suggested that temperature may be the dominant explanation for the spectral sequence seen in SNe Ia around maximum light. The spectra generated in \cite{nugent1995} were synthesised with a model derived from the W7 deflagration model \citep{nomoto1984}. They augmented the Ti abundance by a factor of 10 and were able to reproduce the range of morphologies for the \TiII feature from SN~1991bg through the transitional objects to the normal population simply by altering the temperature, unlike our findings here. The W7 Ti abundance is very compact, dropping off sharply above $\sim$12000~\kms. This results in a model that is very sensitive to temperature changes, as with all of the Ti concentrated around the photosphere, a shift in photospheric temperature will result in an ionisation shift for almost all of the Ti present. The fiducial SN~2021rhu model from \cite{harvey2023} possesses a far more extended Ti distribution - as does the N100 model - and is therefore, far less sensitive to changing photospheric temperatures. Over the range 10500~-~12000~\kms the SN~2021rhu model contains $\sim$11 times as much Ti than the W7 model, similar to that of the model investigated in \cite{nugent1995}.

It is well known that SNe Ia in elliptical early-type galaxies are found to be intrinsically fainter with faster evolving light curves than SNe Ia in spiral late-type galaxies \citep{hamuy1995, gallagher2005, sullivan2006}. While the host galaxy studies are not the scope of this paper, all the `04gs-like' faint normal events in our sample are from late-type, high-mass host galaxies (log$_{10}$ (\Mstar/\Msun) $\gtrsim$ 10) as is also mainly seen for transitional and `91bg-like' events. Hence, for faint `normal' SNe Ia like SN 2004gs, our findings agree with the previous studies showing faint normal SNe Ia preferring to explode in late-type galaxies.

\begin{figure}
\centering
\includegraphics[width=\columnwidth]{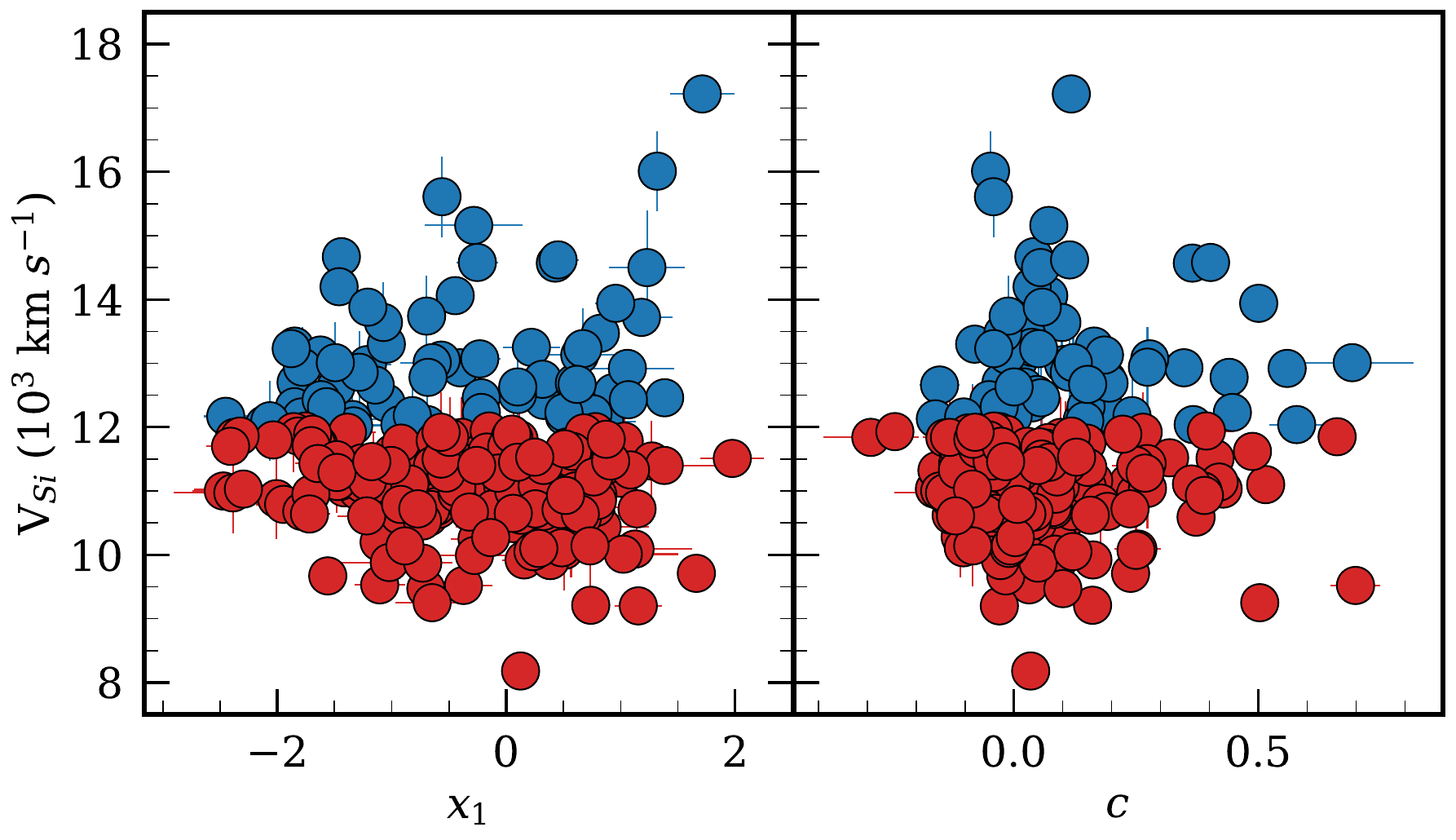}
\caption{The \SiII\ 6355 velocity plotted against SN light curves parameters from SALT2, $x_1$ and $c$. Blue circles represent 66 HV SN Ia and red circles represent 191 NV SN Ia. All SN Ia in this plot are `normal' SNe Ia.}
    \label{fig:x1_c_vel_6355}
\end{figure}

\begin{figure}
\centering
\includegraphics[width=\columnwidth]{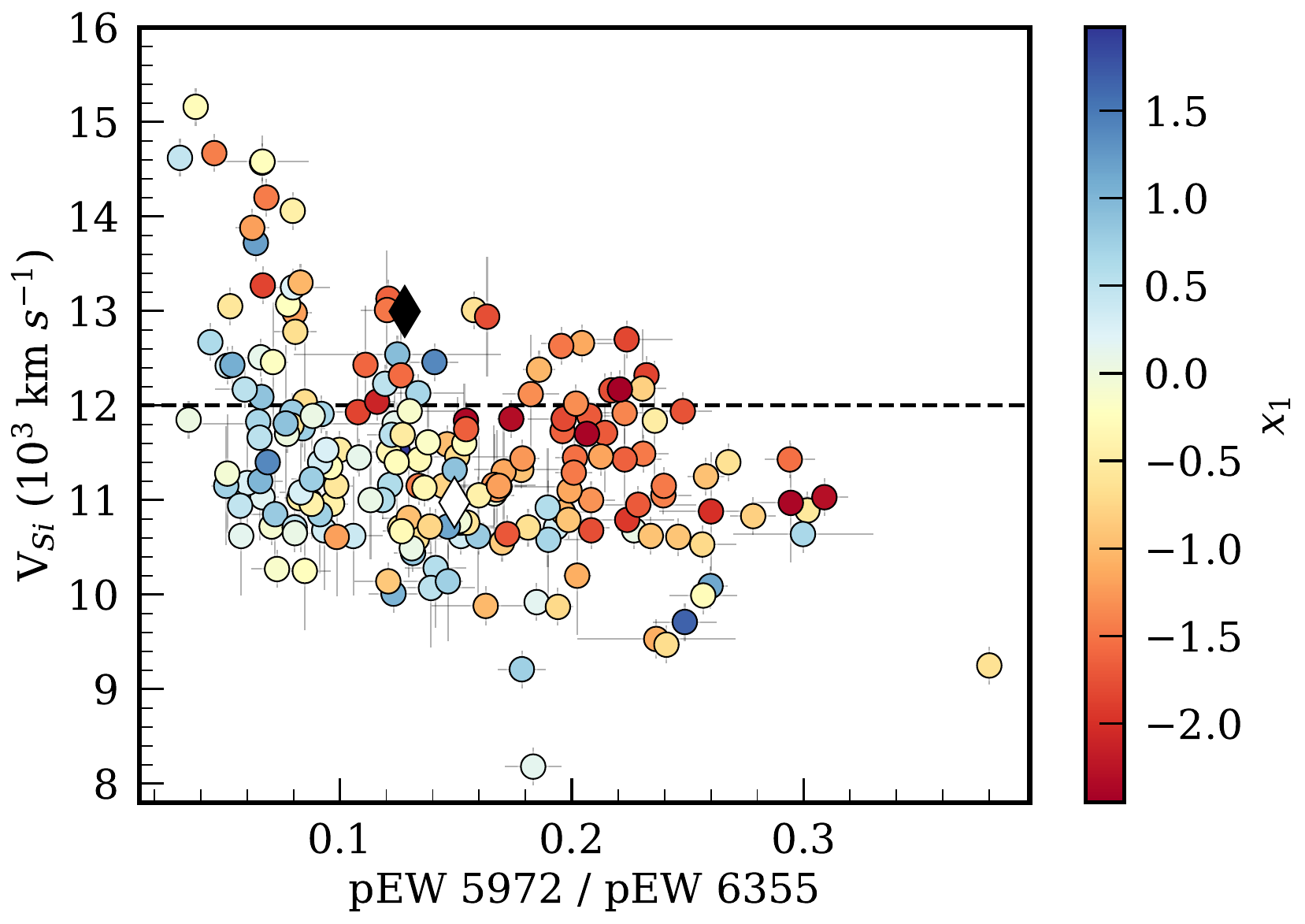}
\caption{\SiIIs\ velocities plotted against the pEW ratio of \SiII\ features, colour mapped with $x_1$ for normal SNe Ia in our sample. The black dashed line at $v_{\mathrm{Si}}$ $=$ 12,000 km s$^{-1}$ marks the boundary between HV and NV SNe Ia on the plot. SNe Ia with velocities higher than 12,000 km s$^{-1}$ are classified as HV SNe Ia, while those below are categorized as NV SNe Ia. Filled and empty black diamond represents the weighted averages of HV SNe Ia and NV SNe Ia, respectively.}
    \label{fig:vsil_pew_ratio_x1}
\end{figure}
\subsection{Connection with light curve properties}
\label{sec:LCprop}

In this section, we compare the spectral line measurements of our SN Ia sample to those of their SALT2 light curve parameters ($x_1$, $c$). We present the peak \SiIIs\ velocity, $v_{\mathrm{Si}}$ as a function of $x_1$ and $c$ for the `normal' SN Ia sample in Fig.~\ref{fig:x1_c_vel_6355}. We adopt the NV/HV classification scheme of \citet{wang2009_jul_b} for splitting the sample based on their velocities. The weighted average of $x_1$ for the HV and NV SNe Ia in our sample are $-$0.68$\pm$0.13 and $-$0.18$\pm$0.07, respectively, showing a 3.4$\sigma$ difference between the two samples. The two populations (NV/HV) show similar distributions with the SALT2 $c$ parameter as seen in \cite{dettman2021}, with mean $c$ for the NV and HV SNe Ia for our sample of $\sim$ 0.04$\pm$0.01 and $\sim$ 0.10$\pm$0.02, respectively, showing a 2.7$\sigma$ difference between the two samples.

We also investigated the relation amongst \SiIIs\ velocity, \SiII\ ratio, and light curve width, $x_1$ for normal SNe Ia (Fig.~\ref{fig:vsil_pew_ratio_x1}). A 3$\sigma$ difference is found with HV SNe Ia, on average, tending to have smaller pEW ratios than NV SNe Ia, with weighted averages of the pEW of 0.117$\pm$0.008 and 0.148$\pm$0.005 for the HV and NV SNe Ia, respectively (shown as diamonds in Fig.~\ref{fig:vsil_pew_ratio_x1}). The \SiII\ ratio is considered an indicator of the temperature of the line-forming region \citep{nugent1995,hachinger2008}. \md{Harvey et al.~(subm.)} estimated that around 40 per cent of SN Ia spectra between $-$5 d and peak have \SiIIs\ high-velocity features that are blended with the photospheric components. While this will not change the photometric properties of the SNe Ia, it will result in larger \SiIIs\ pEWs and higher average velocities for the SNe Ia where a high-velocity feature is present but not distinct from the photospheric component (also see Section \ref{sec:CN_and_BL}). \md{Harvey et al.~(subm.)} did not however investigate the presence of Si HVFs in the \SiIIf\ line and as such we cannot comment upon the resulting impact upon the \SiII\ ratios. md{Harvey et al.~(subm.)} found that high-velocity features in \SiIIs\ were more common in SNe Ia with smaller $x_1$ values. This suggests that the trend we observed, where HV SNe Ia tend to have smaller $x_1$ values compared to NV SNe Ia, may be influenced by contributions from these high-velocity features.

\begin{figure}
\centering
\includegraphics[width=\columnwidth]{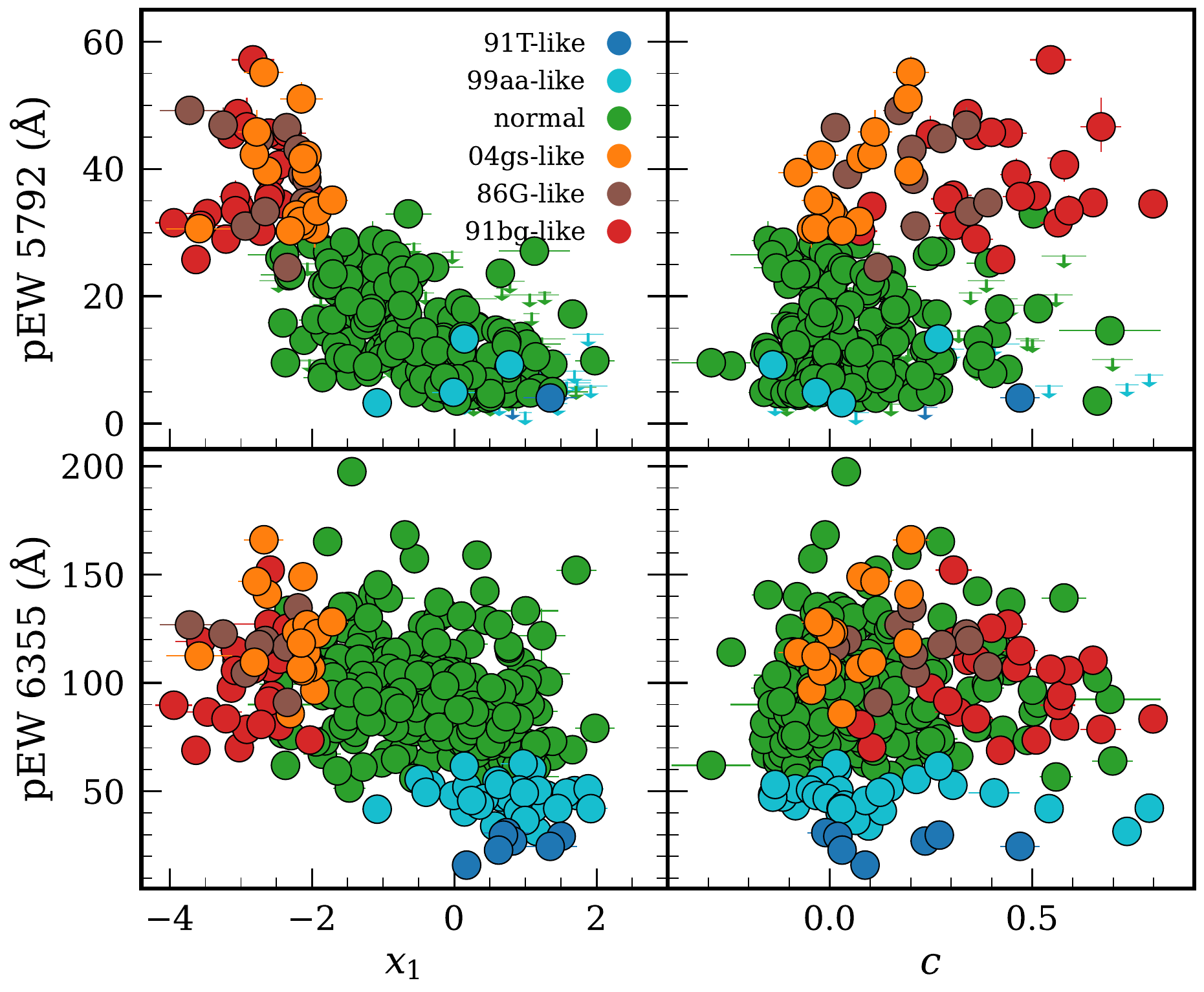}
\caption{\textbf{Top left}: Light curve parameter $x_1$ from SALT2 is plotted against pEW of \SiIIf\ feature. Upper limits represent where actual calculations for \SiIIs\ exist but 5927 does not. \textbf{Top right}: Same plot as left but for the light curve parameter $c$ from SALT2. \textbf{Bottom left}: $x_1$ plotted against pEW of \SiIIs\ feature. \textbf{Bottom right}: Same plot as bottom left but for $c$.}
    \label{fig:x1_R_pEWs}
\end{figure}

We further inspected the light-curve properties of all the objects (not just normal events) in our sample. We compare the pEW of \SiIIf\ and \SiIIs\ with the SALT2 light curve parameters, $x_1$ and $c$ in Fig.~\ref{fig:x1_R_pEWs}. We see a clear relation, similar to that of previous studies such as \cite{nugent1995, hachinger2008}, between  the pEW of the \SiIIf\ features and the light-curve width parameter, $x_1$, where the `91bg-like', `86G-like', and the `cool' normal `04gs-like' events, have on average faster declining light curves ($x_1$ $\lesssim$ $-$2) than `normal' SNe Ia. The `99a-like' and `91T-like' SNe Ia, populate the slower light curve decline rate ($x_1$ $\gtrsim$ 0) regime with similar pEW of \SiIIf\ to normal SNe Ia. Within the `normal' SNe Ia, a trend of SNe Ia with narrower light curves being slightly cooler than those with a broader one is also seen (visible also in Fig.~\ref{fig:vsil_pew_ratio_x1}). This trend is similar to what we already see for `86G-like' and `99aa-like' SNe Ia on the extremes. This trend is now also observed to be present as a gradient within the `normal' SNe Ia.

\begin{figure}
\centering
\includegraphics[width=\columnwidth]{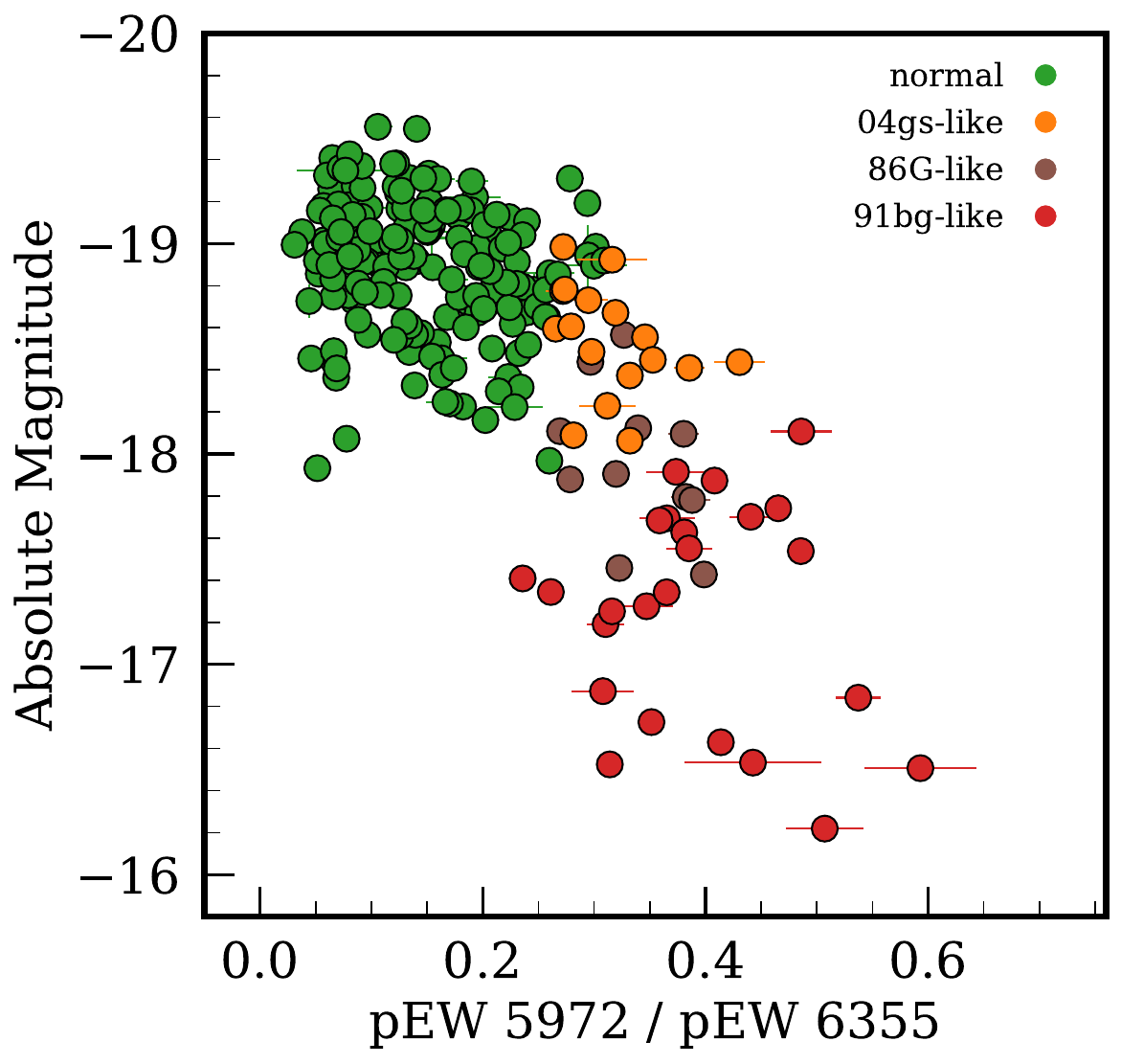}
\caption{Absolute g-band magnitudes of the volume limited ZTF DR2 SN Ia sample from SALT2 as function of \SiII\ ratios. All magnitudes are corrected for MW extinction. In this plot, `normal' SNe Ia with a colour parameter ($c$) higher than 0.3 are excluded to remove highly reddened SNe Ia.}
\label{fig:Rvsabs_g}
\end{figure}

A similar correlation can also be seen between $c$ and \SiIIf\ feature (top right panel of Fig.~\ref{fig:x1_R_pEWs}). The colour $c$, on average is larger for `91bg-like' SNe (weighted average $c$ of 0.50$\pm$0.04 compared to the `86g-like' SNe ($c$ of 0.35$\pm$0.04) and the `04gs-like' SNe Ia ($c$ of 0.04$\pm$0.02). `91T-like' SN Ia are slightly redder ($c$ of 0.14$\pm$0.07) than `99a-like' SN Ia ($c$ of 0.09$\pm$0.04) but the number of `91T-like' SNe Ia is low. No strong correlation between \SiIIs\ and either $x_1$ or $c$ is seen (bottom panel of Fig.~\ref{fig:x1_R_pEWs}), apart from `99aa-like' and `91T-like' events having weaker pEW. This is likely due to be the temperature having a more direct effect on the \SiIIf\ than \SiIIs\ line strengths (see Sec.~\ref{sec:04gs_cont}).
The absolute peak magnitudes from SALT2 (corrected for Milky Way extinction) in the ZTF \textit{g}-band are plotted against the pEW ratio of the \SiII\ 5972 to 6355 \AA\ features in Fig.~\ref{fig:Rvsabs_g}. Only for this plot, we have removed `normal' SNe Ia with $c$ values higher than 0.3 to remove highly reddened normal SNe Ia. `04gs-like' SNe Ia are seen to be on average brighter ($-$18.55$\pm$0.06 mag) than the `86G-like' ($-$17.96$\pm$0.11 mag) and `91bg-like' ($-$17.25$\pm$0.11 mag) events. As shown in Section \ref{sec:04gs_cont} and in previous studies such as \citet{hachinger2008}, the pEW ratio of the \SiII\ features is a good temperature indicator, with the \SiIIf\ feature becoming more prominent when the temperature is lower. The absolute magnitude of an event directly relates to the amount of \Nifs\ produced in the explosion. Therefore, Fig.~\ref{fig:Rvsabs_g} shows that there is a strong relation between the mass of radioactive \Nifs\ and the temperature of the ejecta, with SNe with more \Nifs\ having hotter ejecta. However, from our \textsc{tardis} analysis in Sec.~\ref{sec:04gs_cont}, temperature alone is insufficient to produce the combination of features seen in the spectra of SN 2004gs of weaker \TiII compared to `86G-like' or `91bg-like' events but a similar pEW ratio of the \SiII\ 5972 to 6355 \AA. 

\section{Conclusions}
\label{sec:conclusions}

In this work, we presented a sub-sample of maximum-light spectra of 482 SNe Ia from the ZTF DR2 sample up to a redshift of 0.06. We have investigated the spectral profiles of \SiIIs\ and \SiIIf, as well as tested the template-matching accuracy of the DR2 spectral classifications. We provide a summary of our conclusions:

\begin{enumerate}
      \item We investigated the  accuracy of the spectral-template matching and initial classifications and methods used for typing in 482 SNe Ia. After this manual inspection, we re-classified one SNe Ia with no sub-class (typed as `\uncIa') as `91T-like', 12 `\uncIa' as `99a-like' and 10 `\uncIa' as `normal'. We also re-classified 9 `91T-like' SNe Ia as `99a-like' and 2 `91T-like' SNe Ia as `normal'. Our classifications of `91T-like' and `99aa-like' SNe Ia are consistent with \citet{phillips2022}. While `91T-like' and `99aa-like' SNe Ia are generally included in cosmology samples, there are studies \citep{scalzo2012, yang2022, chakraborty2023} suggesting that inclusion of these event might introduce a systematic bias. It is important to have a more accurate sub-class identification to help keep these cosmological samples of SNe Ia more homogeneous.
      
      \item Host galaxy contamination in the SN Ia spectra is seen to have a considerable impact on sub-typing and spectral features, such as pseudo equivalent widths. No significant change in the velocities is seen. The galaxy contribution makes the SN Ia spectra appear redder and the strength of the features can become weaker. However, the minimum of the feature used to calculate the velocities is less affected.
      \item Spectral template matching, such as \textsc{snid}, still provides fast and reliable results for normal SN Ia. The rarer sub-types, on the other hand, need more detailed analysis and constraints from many spectral features such as \SiII\ ratios, \TiII, \OI, and \CaII NIR. These rarer sub-types are also less represented in training sets for spectral classifications codes. 
      \item Transitional `04gs-like' events have a relatively high \SiII\ ratio (suggesting cooler temperatures than normal SNe Ia) and weak \TiII absorption features. Using \textsc{tardis} modelling, we demonstrated that decreasing the temperature alone produces a good match to the \SII\ ratio but not to the \TiII feature. An increase in the \TiII abundance  is also required to match its spectra. These events appear to sit between normal and sub-luminous SNe Ia but brighter than the standard transitional `86G-like' events.
     \item We assigned Branch sub-types to all SNe in our sample and studied the observed diversity from the spectral features. The `normal' to `91T-like' transition, where `99aa-like' SNe Ia bridging the gap is seen in terms of the pEW  of \SiIIs. True `91T-like' SNe Ia in our sample remain very rare. 
     \item We observed a strong correlation between $x_1$ and the pEW of \SiIIf, with a clear transition in properties at $\sim$30\angstrom\ and $x_1$ of around $-$2. Other than all true `91T-like' SNe Ia showing positive $x_1$ values, no other significant correlation is seen between \SiIIs\ feature and $x_1$ and $c$. Especially for fainter objects, this is due to the \SiIIs\ feature becoming saturated, while the \SiIIf\ feature becomes more prominent due to the ionisation balance. 
     \item We found a trend among \SiIIs\ velocity, \SiII\ ratio and $x_1$ within the `normal' SNe Ia, showing HV SNe Ia with lower pEW ratios tend to have faster declining light curves.
     \item Absolute magnitude of SNe Ia follows a sequence as a function of \SiII\ pEW ratio for the subtypes. By investigating the faint normal SNe Ia, we show that this is likely caused by the different initial abundance profiles and temperature settings, which more directly correlates with the subtypes.
   \end{enumerate}

We emphasize that regardless of the underlying explosion mechanism and/or binary configuration, the continuity seen in both bright and faint SNe Ia persists and is encouraged to be further quantified. Since only the peak spectra are investigated in this work, there is a possibility that even if there are several explosion mechanisms, they may all be showing similar properties due to all being highly driven by \Nifs\ near the peak. Therefore, it is encouraged to obtain more early time spectra to investigate the ejecta properties that are more directly related to the explosion and late time spectra to investigate progenitor channels since the materials at this phase get optically thin. A forthcoming paper will present any effect coming from the host galaxies and their impact on our findings in current cosmological studies.

\begin{acknowledgements}
Based on observations obtained with the Samuel Oschin Telescope 48-inch and the 60-inch Telescope at the Palomar Observatory as part of the Zwicky Transient Facility project. ZTF is supported by the National Science Foundation under Grants No. AST-1440341 and AST-2034437 and a collaboration including current partners Caltech, IPAC, the Weizmann Institute of Science, the Oskar Klein Center at Stockholm University, the University of Maryland, Deutsches Elektronen-Synchrotron and Humboldt University, the TANGO Consortium of Taiwan, the University of Wisconsin at Milwaukee, Trinity College Dublin, Lawrence Livermore National Laboratories, IN2P3, University of Warwick, Ruhr University Bochum, Northwestern University and former partners the University of Washington, Los Alamos National Laboratories, and Lawrence Berkeley National Laboratories. Operations are conducted by COO, IPAC, and UW. 

SED Machine is based upon work supported by the National Science Foundation under Grant No. 1106171. The ZTF forced-photometry service was funded under the Heising-Simons Foundation grant \#12540303 (PI: Graham). This work was supported by the GROWTH project funded by the National Science Foundation under Grant No 1545949 \citep{kasliwal2019}. Fritz \citep{vander2019, coughlin2020} is used in this work. The Gordon and Betty Moore Foundation, through both the Data-Driven Investigator Program and a dedicated grant, provided critical funding for SkyPortal. UB, KM, GD, RS, MD, and JHT are supported by the H2020 European Research Council grant no. 758638. LH is funded by the Irish Research Council under grant number GOIPG/2020/1387. Y.-L.K. has received funding from the Science and Technology Facilities Council [grant number ST/V000713/1]. L.G. acknowledges financial support from the Spanish Ministerio de Ciencia e Innovaci\'on (MCIN) and the Agencia Estatal de Investigaci\'on (AEI) 10.13039/501100011033 under the PID2020-115253GA-I00 HOSTFLOWS project, from Centro Superior de Investigaciones Cient\'ificas (CSIC) under the PIE project 20215AT016 and the program Unidad de Excelencia Mar\'ia de Maeztu CEX2020-001058-M, and from the Departament de Recerca i Universitats de la Generalitat de Catalunya through the 2021-SGR-01270 grant. This project has received funding from the European Research Council (ERC) under the European Union's Horizon 2020 research and innovation programme (grant agreement no 759194 - USNAC). This work has been supported by the research project grant "Understanding the Dynamic Universe" funded by the Knut and Alice Wallenberg Foundation under Dnr KAW 2018.0067. AG acknowledges support from  {\em Vetenskapsr\aa det}, the Swedish Research Council, project 2020-03444. SD acknowledges support from the Marie Curie Individual Fellowship under grant ID 890695 and a Junior Research Fellowship at Lucy Cavendish College. T.E.M.B. acknowledges financial support from the Spanish Ministerio de Ciencia e Innovaci\'on (MCIN), the Agencia Estatal de Investigaci\'on (AEI) 10.13039/501100011033, and the European Union Next Generation EU/PRTR funds under the 2021 Juan de la Cierva program FJC2021-047124-I and the PID2020-115253GA-I00 HOSTFLOWS project, from Centro Superior de Investigaciones Cient\'ificas (CSIC) under the PIE project 20215AT016, and the program Unidad de Excelencia Mar\'ia de Maeztu CEX2020-001058-M. This work has been supported by the Agence Nationale de la Recherche of the French government through the program ANR-21-CE31-0016-03.
 
\end{acknowledgements}

\bibliographystyle{aa}
\bibliography{aanda} 

\begin{thebibliography}{92}
\expandafter\ifx\csname natexlab\endcsname\relax\def\natexlab#1{#1}\fi

\bibitem[{{Ashall} {et~al.}(2016){Ashall}, {Mazzali}, {Pian}, \&
  {James}}]{ashall2016}
{Ashall}, C., {Mazzali}, P.~A., {Pian}, E., \& {James}, P.~A. 2016, \mnras,
  463, 1891

\bibitem[{{Bellm} {et~al.}(2019){Bellm}, {Kulkarni}, {Barlow}, {Feindt},
  {Graham}, {Goobar}, {Kupfer}, {Ngeow}, {Nugent}, {Ofek}, {Prince}, {Riddle},
  {Walters}, \& {Ye}}]{bellm2019}
{Bellm}, E.~C., {Kulkarni}, S.~R., {Barlow}, T., {et~al.} 2019, \pasp, 131,
  068003

\bibitem[{{Benetti} {et~al.}(2005){Benetti}, {Cappellaro}, {Mazzali},
  {Turatto}, {Altavilla}, {Bufano}, {Elias-Rosa}, {Kotak}, {Pignata}, {Salvo},
  \& {Stanishev}}]{benetti2005}
{Benetti}, S., {Cappellaro}, E., {Mazzali}, P.~A., {et~al.} 2005, \apj, 623,
  1011

\bibitem[{{Benetti} {et~al.}(2006){Benetti}, {Cappellaro}, {Turatto},
  {Taubenberger}, {Harutyunyan}, \& {Valenti}}]{Benetti2006}
{Benetti}, S., {Cappellaro}, E., {Turatto}, M., {et~al.} 2006, \apjl, 653, L129

\bibitem[{{Benetti} {et~al.}(2004){Benetti}, {Meikle}, {Stehle}, {Altavilla},
  {Desidera}, {Folatelli}, {Goobar}, {Mattila}, {Mendez}, {Navasardyan},
  {Pastorello}, {Patat}, {Riello}, {Ruiz-Lapuente}, {Tsvetkov}, {Turatto},
  {Mazzali}, \& {Hillebrandt}}]{benetti2004_02bo}
{Benetti}, S., {Meikle}, P., {Stehle}, M., {et~al.} 2004, \mnras, 348, 261

\bibitem[{{Blagorodnova} {et~al.}(2018){Blagorodnova}, {Neill}, {Walters},
  {Kulkarni}, {Fremling}, {Ben-Ami}, {Dekany}, {Fucik}, {Konidaris}, {Nash},
  {Ngeow}, {Ofek}, {O' Sullivan}, {Quimby}, {Ritter}, \&
  {Vyhmeister}}]{blagorodnova2018}
{Blagorodnova}, N., {Neill}, J.~D., {Walters}, R., {et~al.} 2018, \pasp, 130,
  035003

\bibitem[{{Blondin} {et~al.}(2006){Blondin}, {Dessart}, {Leibundgut}, {Branch},
  {H{\"o}flich}, {Tonry}, {Matheson}, {Foley}, {Chornock}, {Filippenko},
  {Sollerman}, {Spyromilio}, {Kirshner}, {Wood-Vasey}, {Clocchiatti},
  {Aguilera}, {Barris}, {Becker}, {Challis}, {Covarrubias}, {Davis},
  {Garnavich}, {Hicken}, {Jha}, {Krisciunas}, {Li}, {Miceli}, {Miknaitis},
  {Pignata}, {Prieto}, {Rest}, {Riess}, {Salvo}, {Schmidt}, {Smith}, {Stubbs},
  \& {Suntzeff}}]{blondin2006}
{Blondin}, S., {Dessart}, L., {Leibundgut}, B., {et~al.} 2006, \aj, 131, 1648

\bibitem[{{Blondin} {et~al.}(2012){Blondin}, {Matheson}, {Kirshner}, {Mandel},
  {Berlind}, {Calkins}, {Challis}, {Garnavich}, {Jha}, {Modjaz}, {Riess}, \&
  {Schmidt}}]{blondin2012}
{Blondin}, S., {Matheson}, T., {Kirshner}, R.~P., {et~al.} 2012, \aj, 143, 126

\bibitem[{{Blondin} \& {Tonry}(2007)}]{Blondin2007}
{Blondin}, S. \& {Tonry}, J.~L. 2007, \apj, 666, 1024

\bibitem[{{Branch} {et~al.}(2009){Branch}, {Chau Dang}, \&
  {Baron}}]{branch2009}
{Branch}, D., {Chau Dang}, L., \& {Baron}, E. 2009, \pasp, 121, 238

\bibitem[{{Branch} {et~al.}(2006){Branch}, {Dang}, {Hall}, {Ketchum},
  {Melakayil}, {Parrent}, {Troxel}, {Casebeer}, {Jeffery}, \&
  {Baron}}]{branch2006}
{Branch}, D., {Dang}, L.~C., {Hall}, N., {et~al.} 2006, \pasp, 118, 560

\bibitem[{{Burgaz} {et~al.}(2021){Burgaz}, {Maeda}, {Kalomeni}, {Kawabata},
  {Yamanaka}, {Kawabata}, {Kawahara}, \& {Nakaoka}}]{burgaz2021}
{Burgaz}, U., {Maeda}, K., {Kalomeni}, B., {et~al.} 2021, \mnras, 502, 4112

\bibitem[{{Burrow} {et~al.}(2020){Burrow}, {Baron}, {Ashall}, {Burns},
  {Morrell}, {Stritzinger}, {Brown}, {Folatelli}, {Freedman}, {Galbany},
  {Hoeflich}, {Hsiao}, {Krisciunas}, {Phillips}, {Piro}, {Suntzeff}, \&
  {Uddin}}]{burrow2020}
{Burrow}, A., {Baron}, E., {Ashall}, C., {et~al.} 2020, \apj, 901, 154

\bibitem[{{Chakraborty} {et~al.}(2023){Chakraborty}, {Sadler}, {Hoeflich},
  {Hsiao}, {Phillips}, {Burns}, {Diamond}, {Dominguez}, {Galbany}, {Uddin},
  {Ashall}, {Krisciunas}, {Kumar}, {Mera}, {Morrell}, {Baron}, {Contreras},
  {Stritzinger}, \& {Suntzeff}}]{chakraborty2023}
{Chakraborty}, S., {Sadler}, B., {Hoeflich}, P., {et~al.} 2023, arXiv e-prints,
  arXiv:2311.03473

\bibitem[{{Chambers} {et~al.}(2016){Chambers}, {Magnier}, {Metcalfe},
  {Flewelling}, {Huber}, {Waters}, {Denneau}, {Draper}, {Farrow}, {Finkbeiner},
  {Holmberg}, {Koppenhoefer}, {Price}, {Rest}, {Saglia}, {Schlafly}, {Smartt},
  {Sweeney}, {Wainscoat}, {Burgett}, {Chastel}, {Grav}, {Heasley}, {Hodapp},
  {Jedicke}, {Kaiser}, {Kudritzki}, {Luppino}, {Lupton}, {Monet}, {Morgan},
  {Onaka}, {Shiao}, {Stubbs}, {Tonry}, {White}, {Ba{\~n}ados}, {Bell},
  {Bender}, {Bernard}, {Boegner}, {Boffi}, {Botticella}, {Calamida},
  {Casertano}, {Chen}, {Chen}, {Cole}, {Deacon}, {Frenk}, {Fitzsimmons},
  {Gezari}, {Gibbs}, {Goessl}, {Goggia}, {Gourgue}, {Goldman}, {Grant},
  {Grebel}, {Hambly}, {Hasinger}, {Heavens}, {Heckman}, {Henderson}, {Henning},
  {Holman}, {Hopp}, {Ip}, {Isani}, {Jackson}, {Keyes}, {Koekemoer}, {Kotak},
  {Le}, {Liska}, {Long}, {Lucey}, {Liu}, {Martin}, {Masci}, {McLean}, {Mindel},
  {Misra}, {Morganson}, {Murphy}, {Obaika}, {Narayan}, {Nieto-Santisteban},
  {Norberg}, {Peacock}, {Pier}, {Postman}, {Primak}, {Rae}, {Rai}, {Riess},
  {Riffeser}, {Rix}, {R{\"o}ser}, {Russel}, {Rutz}, {Schilbach}, {Schultz},
  {Scolnic}, {Strolger}, {Szalay}, {Seitz}, {Small}, {Smith}, {Soderblom},
  {Taylor}, {Thomson}, {Taylor}, {Thakar}, {Thiel}, {Thilker}, {Unger},
  {Urata}, {Valenti}, {Wagner}, {Walder}, {Walter}, {Watters}, {Werner},
  {Wood-Vasey}, \& {Wyse}}]{Chambers2016}
{Chambers}, K.~C., {Magnier}, E.~A., {Metcalfe}, N., {et~al.} 2016, arXiv
  e-prints, arXiv:1612.05560

\bibitem[{{Childress} {et~al.}(2013){Childress}, {Scalzo}, {Sim}, {Tucker},
  {Yuan}, {Schmidt}, {Cenko}, {Silverman}, {Contreras}, {Hsiao}, {Phillips},
  {Morrell}, {Jha}, {McCully}, {Filippenko}, {Anderson}, {Benetti}, {Bufano},
  {de Jaeger}, {Forster}, {Gal-Yam}, {Le Guillou}, {Maguire}, {Maund},
  {Mazzali}, {Pignata}, {Smartt}, {Spyromilio}, {Sullivan}, {Taddia},
  {Valenti}, {Bayliss}, {Bessell}, {Blanc}, {Carson}, {Clubb}, {de Burgh-Day},
  {Desjardins}, {Fang}, {Fox}, {Gates}, {Ho}, {Keller}, {Kelly}, {Lidman},
  {Loaring}, {Mould}, {Owers}, {Ozbilgen}, {Pei}, {Pickering}, {Pracy}, {Rich},
  {Schaefer}, {Scott}, {Stritzinger}, {Vogt}, \& {Zhou}}]{childress2013}
{Childress}, M.~J., {Scalzo}, R.~A., {Sim}, S.~A., {et~al.} 2013, \apj, 770, 29

\bibitem[{{Coughlin} {et~al.}(2023){Coughlin}, {Bloom}, {Nir}, {Antier}, {du
  Laz}, {van der Walt}, {Crellin-Quick}, {Culino}, {Duev}, {Goldstein},
  {Healy}, {Karambelkar}, {Lilleboe}, {Shin}, {Singer}, {Ahumada}, {Anand},
  {Bellm}, {Dekany}, {Graham}, {Kasliwal}, {Kostadinova}, {Kiendrebeogo},
  {Kulkarni}, {Jenkins}, {LeBaron}, {Mahabal}, {Neill}, {Parazin}, {Peloton},
  {Perley}, {Riddle}, {Rusholme}, {van Santen}, {Sollerman}, {Stein}, {Turpin},
  {Wold}, {Amat}, {Bonnefon}, {Bonnefoy}, {Flament}, {Kerkow}, {Kishore},
  {Jani}, {Mahanty}, {Liu}, {Llinares}, {Makarison}, {Olli{\'e}ric}, {Perez},
  {Pont}, \& {Sharma}}]{coughlin2020}
{Coughlin}, M.~W., {Bloom}, J.~S., {Nir}, G., {et~al.} 2023, \apjs, 267, 31

\bibitem[{{Dekany} {et~al.}(2020){Dekany}, {Smith}, {Riddle}, {Feeney},
  {Porter}, {Hale}, {Zolkower}, {Belicki}, {Kaye}, {Henning}, {Walters},
  {Cromer}, {Delacroix}, {Rodriguez}, {Reiley}, {Mao}, {Hover}, {Murphy},
  {Burruss}, {Baker}, {Kowalski}, {Reif}, {Mueller}, {Bellm}, {Graham}, \&
  {Kulkarni}}]{dekany2020}
{Dekany}, R., {Smith}, R.~M., {Riddle}, R., {et~al.} 2020, \pasp, 132, 038001

\bibitem[{{Dettman} {et~al.}(2021){Dettman}, {Jha}, {Dai}, {Foley}, {Rest},
  {Scolnic}, {Siebert}, {Chambers}, {Coulter}, {Huber}, {Johnson}, {Jones},
  {Kilpatrick}, {Kirshner}, {Pan}, {Riess}, \& {Shultz}}]{dettman2021}
{Dettman}, K.~G., {Jha}, S.~W., {Dai}, M., {et~al.} 2021, \apj, 923, 267

\bibitem[{{Dilday} {et~al.}(2012){Dilday}, {Howell}, {Cenko}, {Silverman},
  {Nugent}, {Sullivan}, {Ben-Ami}, {Bildsten}, {Bolte}, {Endl}, {Filippenko},
  {Gnat}, {Horesh}, {Hsiao}, {Kasliwal}, {Kirkman}, {Maguire}, {Marcy},
  {Moore}, {Pan}, {Parrent}, {Podsiadlowski}, {Quimby}, {Sternberg}, {Suzuki},
  {Tytler}, {Xu}, {Bloom}, {Gal-Yam}, {Hook}, {Kulkarni}, {Law}, {Ofek},
  {Polishook}, \& {Poznanski}}]{Dilday2012}
{Dilday}, B., {Howell}, D.~A., {Cenko}, S.~B., {et~al.} 2012, Science, 337, 942

\bibitem[{{Filippenko} {et~al.}(1999){Filippenko}, {Li}, \&
  {Leonard}}]{filippenko1999}
{Filippenko}, A.~V., {Li}, W.~D., \& {Leonard}, D.~C. 1999, \iaucirc, 7108, 2

\bibitem[{{Filippenko} {et~al.}(1992{\natexlab{a}}){Filippenko}, {Richmond},
  {Branch}, {Gaskell}, {Herbst}, {Ford}, {Treffers}, {Matheson}, {Ho}, {Dey},
  {Sargent}, {Small}, \& {van Breugel}}]{filippenko1992a}
{Filippenko}, A.~V., {Richmond}, M.~W., {Branch}, D., {et~al.}
  1992{\natexlab{a}}, \aj, 104, 1543

\bibitem[{{Filippenko} {et~al.}(1992{\natexlab{b}}){Filippenko}, {Richmond},
  {Matheson}, {Shields}, {Burbidge}, {Cohen}, {Dickinson}, {Malkan}, {Nelson},
  {Pietz}, {Schlegel}, {Schmeer}, {Spinrad}, {Steidel}, {Tran}, \&
  {Wren}}]{filippenko1992b}
{Filippenko}, A.~V., {Richmond}, M.~W., {Matheson}, T., {et~al.}
  1992{\natexlab{b}}, \apjl, 384, L15

\bibitem[{{Folatelli} {et~al.}(2013){Folatelli}, {Morrell}, {Phillips},
  {Hsiao}, {Campillay}, {Contreras}, {Castell{\'o}n}, {Hamuy}, {Krzeminski},
  {Roth}, {Stritzinger}, {Burns}, {Freedman}, {Madore}, {Murphy}, {Persson},
  {Prieto}, {Suntzeff}, {Krisciunas}, {Anderson}, {F{\"o}rster}, {Maza},
  {Pignata}, {Rojas}, {Boldt}, {Salgado}, {Wyatt}, {Olivares E.}, {Gal-Yam}, \&
  {Sako}}]{folatelli2013}
{Folatelli}, G., {Morrell}, N., {Phillips}, M.~M., {et~al.} 2013, \apj, 773, 53

\bibitem[{{Foley} \& {Kasen}(2011)}]{foleykasen2011}
{Foley}, R.~J. \& {Kasen}, D. 2011, \apj, 729, 55

\bibitem[{{Gall} {et~al.}(2018){Gall}, {Stritzinger}, {Ashall}, {Baron},
  {Burns}, {Hoeflich}, {Hsiao}, {Mazzali}, {Phillips}, {Filippenko},
  {Anderson}, {Benetti}, {Brown}, {Campillay}, {Challis}, {Contreras}, {Elias
  de la Rosa}, {Folatelli}, {Foley}, {Fraser}, {Holmbo}, {Marion}, {Morrell},
  {Pan}, {Pignata}, {Suntzeff}, {Taddia}, {Torres Robledo}, \&
  {Valenti}}]{Gall2018}
{Gall}, C., {Stritzinger}, M.~D., {Ashall}, C., {et~al.} 2018, \aap, 611, A58

\bibitem[{{Gallagher} {et~al.}(2005){Gallagher}, {Garnavich}, {Berlind},
  {Challis}, {Jha}, \& {Kirshner}}]{gallagher2005}
{Gallagher}, J.~S., {Garnavich}, P.~M., {Berlind}, P., {et~al.} 2005, \apj,
  634, 210

\bibitem[{{Ganeshalingam} {et~al.}(2011){Ganeshalingam}, {Li}, \&
  {Filippenko}}]{ganeshalingam2011}
{Ganeshalingam}, M., {Li}, W., \& {Filippenko}, A.~V. 2011, \mnras, 416, 2607

\bibitem[{{Ganeshalingam} {et~al.}(2012){Ganeshalingam}, {Li}, {Filippenko},
  {Silverman}, {Chornock}, {Foley}, {Matheson}, {Kirshner}, {Milne}, {Calkins},
  \& {Shen}}]{ganeshalingam2012}
{Ganeshalingam}, M., {Li}, W., {Filippenko}, A.~V., {et~al.} 2012, \apj, 751,
  142

\bibitem[{{Graham} {et~al.}(2019){Graham}, {Kulkarni}, {Bellm}, {Adams},
  {Barbarino}, {Blagorodnova}, {Bodewits}, {Bolin}, {Brady}, {Cenko}, {Chang},
  {Coughlin}, {De}, {Eadie}, {Farnham}, {Feindt}, {Franckowiak}, {Fremling},
  {Gezari}, {Ghosh}, {Goldstein}, {Golkhou}, {Goobar}, {Ho}, {Huppenkothen},
  {Ivezi{\'c}}, {Jones}, {Juric}, {Kaplan}, {Kasliwal}, {Kelley}, {Kupfer},
  {Lee}, {Lin}, {Lunnan}, {Mahabal}, {Miller}, {Ngeow}, {Nugent}, {Ofek},
  {Prince}, {Rauch}, {van Roestel}, {Schulze}, {Singer}, {Sollerman}, {Taddia},
  {Yan}, {Ye}, {Yu}, {Barlow}, {Bauer}, {Beck}, {Belicki}, {Biswas}, {Brinnel},
  {Brooke}, {Bue}, {Bulla}, {Burruss}, {Connolly}, {Cromer}, {Cunningham},
  {Dekany}, {Delacroix}, {Desai}, {Duev}, {Feeney}, {Flynn}, {Frederick},
  {Gal-Yam}, {Giomi}, {Groom}, {Hacopians}, {Hale}, {Helou}, {Henning},
  {Hover}, {Hillenbrand}, {Howell}, {Hung}, {Imel}, {Ip}, {Jackson}, {Kaspi},
  {Kaye}, {Kowalski}, {Kramer}, {Kuhn}, {Landry}, {Laher}, {Mao}, {Masci},
  {Monkewitz}, {Murphy}, {Nordin}, {Patterson}, {Penprase}, {Porter},
  {Rebbapragada}, {Reiley}, {Riddle}, {Rigault}, {Rodriguez}, {Rusholme}, {van
  Santen}, {Shupe}, {Smith}, {Soumagnac}, {Stein}, {Surace}, {Szkody}, {Terek},
  {Van Sistine}, {van Velzen}, {Vestrand}, {Walters}, {Ward}, {Zhang}, \&
  {Zolkower}}]{graham2019}
{Graham}, M.~J., {Kulkarni}, S.~R., {Bellm}, E.~C., {et~al.} 2019, \pasp, 131,
  078001

\bibitem[{{Gupta} {et~al.}(2016){Gupta}, {Kuhlmann}, {Kovacs}, {Spinka},
  {Kessler}, {Goldstein}, {Liotine}, {Pomian}, {D'Andrea}, {Sullivan},
  {Carretero}, {Castander}, {Nichol}, {Finley}, {Fischer}, {Foley}, {Kim},
  {Papadopoulos}, {Sako}, {Scolnic}, {Smith}, {Tucker}, {Uddin}, {Wolf},
  {Yuan}, {Abbott}, {Abdalla}, {Benoit-L{\'e}vy}, {Bertin}, {Brooks}, {Carnero
  Rosell}, {Carrasco Kind}, {Cunha}, {da Costa}, {Desai}, {Doel}, {Eifler},
  {Evrard}, {Flaugher}, {Fosalba}, {Gazta{\~n}aga}, {Gruen}, {Gruendl},
  {James}, {Kuehn}, {Kuropatkin}, {Maia}, {Marshall}, {Miquel}, {Plazas},
  {Romer}, {S{\'a}nchez}, {Schubnell}, {Sevilla-Noarbe}, {Sobreira}, {Suchyta},
  {Swanson}, {Tarle}, {Walker}, \& {Wester}}]{gupta2016}
{Gupta}, R.~R., {Kuhlmann}, S., {Kovacs}, E., {et~al.} 2016, \aj, 152, 154

\bibitem[{{Guy} {et~al.}(2007){Guy}, {Astier}, {Baumont}, {Hardin}, {Pain},
  {Regnault}, {Basa}, {Carlberg}, {Conley}, {Fabbro}, {Fouchez}, {Hook},
  {Howell}, {Perrett}, {Pritchet}, {Rich}, {Sullivan}, {Antilogus}, {Aubourg},
  {Bazin}, {Bronder}, {Filiol}, {Palanque-Delabrouille}, {Ripoche}, \&
  {Ruhlmann-Kleider}}]{guy2007}
{Guy}, J., {Astier}, P., {Baumont}, S., {et~al.} 2007, \aap, 466, 11

\bibitem[{{Hachinger} {et~al.}(2008){Hachinger}, {Mazzali}, {Tanaka},
  {Hillebrandt}, \& {Benetti}}]{hachinger2008}
{Hachinger}, S., {Mazzali}, P.~A., {Tanaka}, M., {Hillebrandt}, W., \&
  {Benetti}, S. 2008, \mnras, 389, 1087

\bibitem[{{Hamuy} {et~al.}(2006){Hamuy}, {Folatelli}, {Morrell}, {Phillips},
  {Suntzeff}, {Persson}, {Roth}, {Gonzalez}, {Krzeminski}, {Contreras},
  {Freedman}, {Murphy}, {Madore}, {Wyatt}, {Maza}, {Filippenko}, {Li}, \&
  {Pinto}}]{hamuy2006}
{Hamuy}, M., {Folatelli}, G., {Morrell}, N.~I., {et~al.} 2006, \pasp, 118, 2

\bibitem[{{Hamuy} {et~al.}(1995){Hamuy}, {Phillips}, {Maza}, {Suntzeff},
  {Schommer}, \& {Aviles}}]{hamuy1995}
{Hamuy}, M., {Phillips}, M.~M., {Maza}, J., {et~al.} 1995, \aj, 109, 1

\bibitem[{{Harvey} {et~al.}(2023){Harvey}, {Maguire}, {Magee}, {Bulla},
  {Dhawan}, {Schulze}, {Sollerman}, {Deckers}, {Dimitriadis}, {Reusch},
  {Smith}, {Terwel}, {Coughlin}, {Masci}, {Purdum}, {Reedy}, {Robert}, \&
  {Wold}}]{harvey2023}
{Harvey}, L., {Maguire}, K., {Magee}, M.~R., {et~al.} 2023, \mnras, 522, 4444

\bibitem[{{Hatano} {et~al.}(1999){Hatano}, {Branch}, {Fisher}, {Baron}, \&
  {Filippenko}}]{hatano1999}
{Hatano}, K., {Branch}, D., {Fisher}, A., {Baron}, E., \& {Filippenko}, A.~V.
  1999, \apj, 525, 881

\bibitem[{{Howell} {et~al.}(2006){Howell}, {Sullivan}, {Nugent}, {Ellis},
  {Conley}, {Le Borgne}, {Carlberg}, {Guy}, {Balam}, {Basa}, {Fouchez}, {Hook},
  {Hsiao}, {Neill}, {Pain}, {Perrett}, \& {Pritchet}}]{Howell2006}
{Howell}, D.~A., {Sullivan}, M., {Nugent}, P.~E., {et~al.} 2006, \nat, 443, 308

\bibitem[{{Hoyle} \& {Fowler}(1960)}]{hoyle1960}
{Hoyle}, F. \& {Fowler}, W.~A. 1960, \apj, 132, 565

\bibitem[{{Hsiao} {et~al.}(2015){Hsiao}, {Burns}, {Contreras}, {H{\"o}flich},
  {Sand}, {Marion}, {Phillips}, {Stritzinger}, {Gonz{\'a}lez-Gait{\'a}n},
  {Mason}, {Folatelli}, {Parent}, {Gall}, {Amanullah}, {Anupama}, {Arcavi},
  {Banerjee}, {Beletsky}, {Blanc}, {Bloom}, {Brown}, {Campillay}, {Cao}, {De
  Cia}, {Diamond}, {Freedman}, {Gonzalez}, {Goobar}, {Holmbo}, {Howell},
  {Johansson}, {Kasliwal}, {Kirshner}, {Krisciunas}, {Kulkarni}, {Maguire},
  {Milne}, {Morrell}, {Nugent}, {Ofek}, {Osip}, {Palunas}, {Perley}, {Persson},
  {Piro}, {Rabus}, {Roth}, {Schiefelbein}, {Srivastav}, {Sullivan}, {Suntzeff},
  {Surace}, {Wo{\'z}niak}, \& {Yaron}}]{Hsiao2015}
{Hsiao}, E.~Y., {Burns}, C.~R., {Contreras}, C., {et~al.} 2015, \aap, 578, A9

\bibitem[{{Iben} \& {Tutukov}(1984)}]{iben1984}
{Iben}, I., J. \& {Tutukov}, A.~V. 1984, \apjs, 54, 335

\bibitem[{{Kasliwal} {et~al.}(2019){Kasliwal}, {Cannella}, {Bagdasaryan},
  {Hung}, {Feindt}, {Singer}, {Coughlin}, {Fremling}, {Walters}, {Duev},
  {Itoh}, \& {Quimby}}]{kasliwal2019}
{Kasliwal}, M.~M., {Cannella}, C., {Bagdasaryan}, A., {et~al.} 2019, \pasp,
  131, 038003

\bibitem[{{Kerzendorf} {et~al.}(2020){Kerzendorf}, {Sim}, {Vogl}, {Williamson},
  {P{\'a}ssaro}, {Fl{\"o}rs}, {Camacho}, {Jan{\v{c}}auskas}, {Harpole},
  {N{\"o}bauer}, {Lietzau}, {Mishin}, {Tsamis}, {Boyle}, {Shingles}, {Gupta},
  {Desai}, {Klauser}, {Beaujean}, {Suban-Loewen}, {Heringer}, {Barna},
  {Gautam}, {Barbosa}, {Patel}, {Varanasi}, {Eweis}, {Reinecke}, {Bylund},
  {Bentil}, {Eguren}, {Livneh}, {Singhal}, {O'Brien}, {Rajagopalan}, {Jain},
  {Reichenbach}, {Mishra}, {Singh}, {Sofiatti}, {Selsing}, {Kowalski}, {Savel},
  {Talegaonkar}, {Patel}, {Patra}, {Nayak}, {Kumar}, {Sarafina}, {Gillanders},
  {Sharma}, {Wahi}, {Dasgupta}, {Magee}, {Yap}, \& {Gupta}}]{Kerzendorf2020}
{Kerzendorf}, W., {Sim}, S., {Vogl}, C., {et~al.} 2020, {tardis-sn/tardis:
  TARDIS v3.0.dev3459}, Zenodo

\bibitem[{{Kerzendorf} \& {Sim}(2014)}]{Kerzendorf2014}
{Kerzendorf}, W.~E. \& {Sim}, S.~A. 2014, \mnras, 440, 387

\bibitem[{{Khokhlov}(1991)}]{khokhlov1991}
{Khokhlov}, A.~M. 1991, \aap, 245, 114

\bibitem[{{Kim} {et~al.}(2022){Kim}, {Rigault}, {Neill}, {Briday}, {Copin},
  {Lezmy}, {Nicolas}, {Riddle}, {Sharma}, {Smith}, {Sollerman}, \&
  {Walters}}]{kim2022}
{Kim}, Y.~L., {Rigault}, M., {Neill}, J.~D., {et~al.} 2022, \pasp, 134, 024505

\bibitem[{{Kinney} {et~al.}(1996){Kinney}, {Calzetti}, {Bohlin}, {McQuade},
  {Storchi-Bergmann}, \& {Schmitt}}]{kinney1996}
{Kinney}, A.~L., {Calzetti}, D., {Bohlin}, R.~C., {et~al.} 1996, \apj, 467, 38

\bibitem[{{Li} {et~al.}(2003){Li}, {Filippenko}, {Chornock}, {Berger},
  {Berlind}, {Calkins}, {Challis}, {Fassnacht}, {Jha}, {Kirshner}, {Matheson},
  {Sargent}, {Simcoe}, {Smith}, \& {Squires}}]{li2003}
{Li}, W., {Filippenko}, A.~V., {Chornock}, R., {et~al.} 2003, \pasp, 115, 453

\bibitem[{{Maeda} {et~al.}(2010){Maeda}, {Benetti}, {Stritzinger}, {R{\"o}pke},
  {Folatelli}, {Sollerman}, {Taubenberger}, {Nomoto}, {Leloudas}, {Hamuy},
  {Tanaka}, {Mazzali}, \& {Elias-Rosa}}]{maeda2010}
{Maeda}, K., {Benetti}, S., {Stritzinger}, M., {et~al.} 2010, \nat, 466, 82

\bibitem[{{Maeda} \& {Terada}(2016)}]{maeda2016}
{Maeda}, K. \& {Terada}, Y. 2016, International Journal of Modern Physics D,
  25, 1630024

\bibitem[{{Magee} {et~al.}(2016){Magee}, {Kotak}, {Sim}, {Kromer},
  {Rabinowitz}, {Smartt}, {Baltay}, {Campbell}, {Chen}, {Fink}, {Gal-Yam},
  {Galbany}, {Hillebrandt}, {Inserra}, {Kankare}, {Le Guillou}, {Lyman},
  {Maguire}, {Pakmor}, {R{\"o}pke}, {Ruiter}, {Seitenzahl}, {Sullivan},
  {Valenti}, \& {Young}}]{magee2016}
{Magee}, M.~R., {Kotak}, R., {Sim}, S.~A., {et~al.} 2016, \aap, 589, A89

\bibitem[{{Maguire} {et~al.}(2014){Maguire}, {Sullivan}, {Pan}, {Gal-Yam},
  {Hook}, {Howell}, {Nugent}, {Mazzali}, {Chotard}, {Clubb}, {Filippenko},
  {Kasliwal}, {Kandrashoff}, {Poznanski}, {Saunders}, {Silverman}, {Walker}, \&
  {Xu}}]{maguire2014}
{Maguire}, K., {Sullivan}, M., {Pan}, Y.~C., {et~al.} 2014, \mnras, 444, 3258

\bibitem[{{Maoz} {et~al.}(2014){Maoz}, {Mannucci}, \& {Nelemans}}]{maoz2014}
{Maoz}, D., {Mannucci}, F., \& {Nelemans}, G. 2014, \araa, 52, 107

\bibitem[{{Masci} {et~al.}(2019){Masci}, {Laher}, {Rusholme}, {Shupe}, {Groom},
  {Surace}, {Jackson}, {Monkewitz}, {Beck}, {Flynn}, {Terek}, {Landry},
  {Hacopians}, {Desai}, {Howell}, {Brooke}, {Imel}, {Wachter}, {Ye}, {Lin},
  {Cenko}, {Cunningham}, {Rebbapragada}, {Bue}, {Miller}, {Mahabal}, {Bellm},
  {Patterson}, {Juri{\'c}}, {Golkhou}, {Ofek}, {Walters}, {Graham}, {Kasliwal},
  {Dekany}, {Kupfer}, {Burdge}, {Cannella}, {Barlow}, {Van Sistine}, {Giomi},
  {Fremling}, {Blagorodnova}, {Levitan}, {Riddle}, {Smith}, {Helou}, {Prince},
  \& {Kulkarni}}]{masci2019}
{Masci}, F.~J., {Laher}, R.~R., {Rusholme}, B., {et~al.} 2019, \pasp, 131,
  018003

\bibitem[{{Morrell} {et~al.}(2004){Morrell}, {Folatelli}, \&
  {Hamuy}}]{morell2004}
{Morrell}, N., {Folatelli}, G., \& {Hamuy}, M. 2004, \iaucirc, 8453, 2

\bibitem[{{Nomoto} {et~al.}(1997){Nomoto}, {Iwamoto}, \&
  {Kishimoto}}]{nomoto1997}
{Nomoto}, K., {Iwamoto}, K., \& {Kishimoto}, N. 1997, Science, 276, 1378

\bibitem[{{Nomoto} {et~al.}(1984){Nomoto}, {Thielemann}, \&
  {Yokoi}}]{nomoto1984}
{Nomoto}, K., {Thielemann}, F.~K., \& {Yokoi}, K. 1984, \apj, 286, 644

\bibitem[{{Nordin} {et~al.}(2011){Nordin}, {{\"O}stman}, {Goobar}, {Amanullah},
  {Nichol}, {Smith}, {Sollerman}, {Bassett}, {Frieman}, {Garnavich},
  {Leloudas}, {Sako}, \& {Schneider}}]{nordin2011}
{Nordin}, J., {{\"O}stman}, L., {Goobar}, A., {et~al.} 2011, \aap, 526, A119

\bibitem[{{Nugent} {et~al.}(1995){Nugent}, {Phillips}, {Baron}, {Branch}, \&
  {Hauschildt}}]{nugent1995}
{Nugent}, P., {Phillips}, M., {Baron}, E., {Branch}, D., \& {Hauschildt}, P.
  1995, \apjl, 455, L147

\bibitem[{{Nugent} {et~al.}(2011){Nugent}, {Sullivan}, {Cenko}, {Thomas},
  {Kasen}, {Howell}, {Bersier}, {Bloom}, {Kulkarni}, {Kandrashoff},
  {Filippenko}, {Silverman}, {Marcy}, {Howard}, {Isaacson}, {Maguire},
  {Suzuki}, {Tarlton}, {Pan}, {Bildsten}, {Fulton}, {Parrent}, {Sand},
  {Podsiadlowski}, {Bianco}, {Dilday}, {Graham}, {Lyman}, {James}, {Kasliwal},
  {Law}, {Quimby}, {Hook}, {Walker}, {Mazzali}, {Pian}, {Ofek}, {Gal-Yam}, \&
  {Poznanski}}]{nugent2011}
{Nugent}, P.~E., {Sullivan}, M., {Cenko}, S.~B., {et~al.} 2011, \nat, 480, 344

\bibitem[{{O'Brien} {et~al.}(2023){O'Brien}, {Kerzendorf}, {Fullard}, {Pakmor},
  {Buchner}, {Vogl}, {Chen}, {van der Smagt}, {Williamson}, \&
  {Singhal}}]{obrien2023}
{O'Brien}, J.~T., {Kerzendorf}, W.~E., {Fullard}, A., {et~al.} 2023, arXiv
  e-prints, arXiv:2306.08137

\bibitem[{{O'Brien} {et~al.}(2021){O'Brien}, {Kerzendorf}, {Fullard},
  {Williamson}, {Pakmor}, {Buchner}, {Hachinger}, {Vogl}, {Gillanders},
  {Fl{\"o}rs}, \& {van der Smagt}}]{obrien2021}
{O'Brien}, J.~T., {Kerzendorf}, W.~E., {Fullard}, A., {et~al.} 2021, \apjl,
  916, L14

\bibitem[{{Ogawa} {et~al.}(2023){Ogawa}, {Maeda}, \& {Kawabata}}]{ogawa2023}
{Ogawa}, M., {Maeda}, K., \& {Kawabata}, M. 2023, \apj, 955, 49

\bibitem[{{Park} \& {Li}(2004)}]{park2004}
{Park}, S. \& {Li}, W. 2004, \iaucirc, 8453, 1

\bibitem[{{Perlmutter} {et~al.}(1999){Perlmutter}, {Aldering}, {Goldhaber},
  {Knop}, {Nugent}, {Castro}, {Deustua}, {Fabbro}, {Goobar}, {Groom}, {Hook},
  {Kim}, {Kim}, {Lee}, {Nunes}, {Pain}, {Pennypacker}, {Quimby}, {Lidman},
  {Ellis}, {Irwin}, {McMahon}, {Ruiz-Lapuente}, {Walton}, {Schaefer}, {Boyle},
  {Filippenko}, {Matheson}, {Fruchter}, {Panagia}, {Newberg}, {Couch}, \&
  {Project}}]{perlmutter1999}
{Perlmutter}, S., {Aldering}, G., {Goldhaber}, G., {et~al.} 1999, \apj, 517,
  565

\bibitem[{{Perlmutter} {et~al.}(1997){Perlmutter}, {Gabi}, {Goldhaber},
  {Goobar}, {Groom}, {Hook}, {Kim}, {Kim}, {Lee}, {Pain}, {Pennypacker},
  {Small}, {Ellis}, {McMahon}, {Boyle}, {Bunclark}, {Carter}, {Irwin},
  {Glazebrook}, {Newberg}, {Filippenko}, {Matheson}, {Dopita}, \&
  {Couch}}]{perlmutter1997}
{Perlmutter}, S., {Gabi}, S., {Goldhaber}, G., {et~al.} 1997, \apj, 483, 565

\bibitem[{{Phillips} {et~al.}(2022){Phillips}, {Ashall}, {Burns}, {Contreras},
  {Galbany}, {Hoeflich}, {Hsiao}, {Morrell}, {Nugent}, {Uddin}, {Baron},
  {Freedman}, {Harris}, {Krisciunas}, {Kumar}, {Lu}, {Persson}, {Piro},
  {Polin}, {Shahbandeh}, {Stritzinger}, \& {Suntzeff}}]{phillips2022}
{Phillips}, M.~M., {Ashall}, C., {Burns}, C.~R., {et~al.} 2022, \apj, 938, 47

\bibitem[{{Phillips} {et~al.}(1987){Phillips}, {Phillips}, {Heathcote},
  {Blanco}, {Geisler}, {Hamilton}, {Suntzeff}, {Jablonski}, {Steiner},
  {Cowley}, {Schmidtke}, {Wyckoff}, {Hutchings}, {Tonry}, {Strauss},
  {Thorstensen}, {Honey}, {Maza}, {Ruiz}, {Landolt}, {Uomoto}, {Rich},
  {Grindlay}, {Cohn}, {Smith}, {Lutz}, {Lavery}, \& {Saha}}]{phillips1987}
{Phillips}, M.~M., {Phillips}, A.~C., {Heathcote}, S.~R., {et~al.} 1987, \pasp,
  99, 592

\bibitem[{{Phillips} {et~al.}(1992){Phillips}, {Wells}, {Suntzeff}, {Hamuy},
  {Leibundgut}, {Kirshner}, \& {Foltz}}]{phillips1992}
{Phillips}, M.~M., {Wells}, L.~A., {Suntzeff}, N.~B., {et~al.} 1992, \aj, 103,
  1632

\bibitem[{{Pignata} {et~al.}(2008){Pignata}, {Benetti}, {Mazzali}, {Kotak},
  {Patat}, {Meikle}, {Stehle}, {Leibundgut}, {Suntzeff}, {Buson}, {Cappellaro},
  {Clocchiatti}, {Hamuy}, {Maza}, {Mendez}, {Ruiz-Lapuente}, {Salvo},
  {Schmidt}, {Turatto}, \& {Hillebrandt}}]{pignata2008}
{Pignata}, G., {Benetti}, S., {Mazzali}, P.~A., {et~al.} 2008, \mnras, 388, 971

\bibitem[{{Quimby} {et~al.}(2006){Quimby}, {H{\"o}flich}, {Kannappan},
  {Rykoff}, {Rujopakarn}, {Akerlof}, {Gerardy}, \& {Wheeler}}]{quimby2006}
{Quimby}, R., {H{\"o}flich}, P., {Kannappan}, S.~J., {et~al.} 2006, \apj, 636,
  400

\bibitem[{{Riess} {et~al.}(1998){Riess}, {Filippenko}, {Challis},
  {Clocchiatti}, {Diercks}, {Garnavich}, {Gilliland}, {Hogan}, {Jha},
  {Kirshner}, {Leibundgut}, {Phillips}, {Reiss}, {Schmidt}, {Schommer},
  {Smith}, {Spyromilio}, {Stubbs}, {Suntzeff}, \& {Tonry}}]{ries1998}
{Riess}, A.~G., {Filippenko}, A.~V., {Challis}, P., {et~al.} 1998, \aj, 116,
  1009

\bibitem[{{Riess} {et~al.}(2016){Riess}, {Macri}, {Hoffmann}, {Scolnic},
  {Casertano}, {Filippenko}, {Tucker}, {Reid}, {Jones}, {Silverman},
  {Chornock}, {Challis}, {Yuan}, {Brown}, \& {Foley}}]{ries2016}
{Riess}, A.~G., {Macri}, L.~M., {Hoffmann}, S.~L., {et~al.} 2016, \apj, 826, 56

\bibitem[{{Rigault} {et~al.}(2019){Rigault}, {Neill}, {Blagorodnova}, {Dugas},
  {Feeney}, {Walters}, {Brinnel}, {Copin}, {Fremling}, {Nordin}, \&
  {Sollerman}}]{rigault2019}
{Rigault}, M., {Neill}, J.~D., {Blagorodnova}, N., {et~al.} 2019, \aap, 627,
  A115

\bibitem[{{R{\"o}pke} {et~al.}(2012){R{\"o}pke}, {Kromer}, {Seitenzahl},
  {Pakmor}, {Sim}, {Taubenberger}, {Ciaraldi-Schoolmann}, {Hillebrandt},
  {Aldering}, {Antilogus}, {Baltay}, {Benitez-Herrera}, {Bongard}, {Buton},
  {Canto}, {Cellier-Holzem}, {Childress}, {Chotard}, {Copin}, {Fakhouri},
  {Fink}, {Fouchez}, {Gangler}, {Guy}, {Hachinger}, {Hsiao}, {Chen},
  {Kerschhaggl}, {Kowalski}, {Nugent}, {Paech}, {Pain}, {Pecontal}, {Pereira},
  {Perlmutter}, {Rabinowitz}, {Rigault}, {Runge}, {Saunders}, {Smadja},
  {Suzuki}, {Tao}, {Thomas}, {Tilquin}, \& {Wu}}]{ropke2012}
{R{\"o}pke}, F.~K., {Kromer}, M., {Seitenzahl}, I.~R., {et~al.} 2012, \apjl,
  750, L19

\bibitem[{{Scalzo} {et~al.}(2012){Scalzo}, {Aldering}, {Antilogus}, {Aragon},
  {Bailey}, {Baltay}, {Bongard}, {Buton}, {Canto}, {Cellier-Holzem},
  {Childress}, {Chotard}, {Copin}, {Fakhouri}, {Gangler}, {Guy}, {Hsiao},
  {Kerschhaggl}, {Kowalski}, {Nugent}, {Paech}, {Pain}, {Pecontal}, {Pereira},
  {Perlmutter}, {Rabinowitz}, {Rigault}, {Runge}, {Smadja}, {Tao}, {Thomas},
  {Weaver}, {Wu}, \& {Nearby Supernova Factory}}]{scalzo2012}
{Scalzo}, R., {Aldering}, G., {Antilogus}, P., {et~al.} 2012, \apj, 757, 12

\bibitem[{{Schlafly} \& {Finkbeiner}(2011)}]{schlafly2011}
{Schlafly}, E.~F. \& {Finkbeiner}, D.~P. 2011, \apj, 737, 103

\bibitem[{{Seitenzahl} {et~al.}(2013){Seitenzahl}, {Ciaraldi-Schoolmann},
  {R{\"o}pke}, {Fink}, {Hillebrandt}, {Kromer}, {Pakmor}, {Ruiter}, {Sim}, \&
  {Taubenberger}}]{seitenzahl2013}
{Seitenzahl}, I.~R., {Ciaraldi-Schoolmann}, F., {R{\"o}pke}, F.~K., {et~al.}
  2013, \mnras, 429, 1156

\bibitem[{{Smith} {et~al.}(2012){Smith}, {Nichol}, {Dilday}, {Marriner},
  {Kessler}, {Bassett}, {Cinabro}, {Frieman}, {Garnavich}, {Jha}, {Lampeitl},
  {Sako}, {Schneider}, \& {Sollerman}}]{smith2012}
{Smith}, M., {Nichol}, R.~C., {Dilday}, B., {et~al.} 2012, \apj, 755, 61

\bibitem[{{Soumagnac} {et~al.}(2024){Soumagnac}, {Nugent}, {Knop}, {Ho},
  {Hohensee}, {Awbrey}, {Andersen}, {Aldering}, {Ventura}, {Aguilar}, {Ahlen},
  {Benzvi}, {Brooks}, {Brout}, {Claybaugh}, {Davis}, {Dawson}, {de la Macorra},
  {Dey}, {Dey}, {Doel}, {Douglass}, {Forero-Romero}, {Gaztanaga}, {Gontcho},
  {Graur}, {Guy}, {Hahn}, {Honscheid}, {Howlett}, {Kim}, {Kisner}, {Kremin},
  {Lambert}, {Landriau}, {Lang}, {Le Guillou}, {Manera}, {Meisner}, {Miquel},
  {Moustakas}, {Myers}, {Nie}, {Palmese}, {Parkinson}, {Poppett}, {Prada},
  {Qin}, {Rezaie}, {Rossi}, {Sanchez}, {Schlegel}, {Schubnell}, {Silber},
  {Tarle}, {Weaver}, \& {Zhou}}]{MOST_Hosts}
{Soumagnac}, M.~T., {Nugent}, P., {Knop}, R.~A., {et~al.} 2024, arXiv e-prints,
  arXiv:2405.03857

\bibitem[{{Srivastav} {et~al.}(2020){Srivastav}, {Smartt}, {Leloudas}, {Huber},
  {Chambers}, {Malesani}, {Hjorth}, {Gillanders}, {Schultz}, {Sim}, {Auchettl},
  {Fynbo}, {Gall}, {McBrien}, {Rest}, {Smith}, {Wojtak}, \&
  {Young}}]{srivastav2020}
{Srivastav}, S., {Smartt}, S.~J., {Leloudas}, G., {et~al.} 2020, \apjl, 892,
  L24

\bibitem[{{Stehle} {et~al.}(2005){Stehle}, {Mazzali}, {Benetti}, \&
  {Hillebrandt}}]{stehle2005_02bo}
{Stehle}, M., {Mazzali}, P.~A., {Benetti}, S., \& {Hillebrandt}, W. 2005,
  \mnras, 360, 1231

\bibitem[{{Sullivan} {et~al.}(2006){Sullivan}, {Le Borgne}, {Pritchet},
  {Hodsman}, {Neill}, {Howell}, {Carlberg}, {Astier}, {Aubourg}, {Balam},
  {Basa}, {Conley}, {Fabbro}, {Fouchez}, {Guy}, {Hook}, {Pain},
  {Palanque-Delabrouille}, {Perrett}, {Regnault}, {Rich}, {Taillet}, {Baumont},
  {Bronder}, {Ellis}, {Filiol}, {Lusset}, {Perlmutter}, {Ripoche}, \&
  {Tao}}]{sullivan2006}
{Sullivan}, M., {Le Borgne}, D., {Pritchet}, C.~J., {et~al.} 2006, \apj, 648,
  868

\bibitem[{{van der Walt} {et~al.}(2019){van der Walt}, {Crellin-Quick}, \&
  {Bloom}}]{vander2019}
{van der Walt}, S., {Crellin-Quick}, A., \& {Bloom}, J. 2019, The Journal of
  Open Source Software, 4, 1247

\bibitem[{{Wang} {et~al.}(2009{\natexlab{a}}){Wang}, {Filippenko},
  {Ganeshalingam}, {Li}, {Silverman}, {Wang}, {Chornock}, {Foley}, {Gates},
  {Macomber}, {Serduke}, {Steele}, \& {Wong}}]{wang2009_jul_b}
{Wang}, X., {Filippenko}, A.~V., {Ganeshalingam}, M., {et~al.}
  2009{\natexlab{a}}, \apjl, 699, L139

\bibitem[{{Wang} {et~al.}(2009{\natexlab{b}}){Wang}, {Li}, {Filippenko},
  {Foley}, {Kirshner}, {Modjaz}, {Bloom}, {Brown}, {Carter}, {Friedman},
  {Gal-Yam}, {Ganeshalingam}, {Hicken}, {Krisciunas}, {Milne}, {Silverman},
  {Suntzeff}, {Wood-Vasey}, {Cenko}, {Challis}, {Fox}, {Kirkman}, {Li}, {Li},
  {Malkan}, {Moore}, {Reitzel}, {Rich}, {Serduke}, {Shang}, {Steele}, {Swift},
  {Tao}, {Wong}, \& {Zhang}}]{wang2009_may_a}
{Wang}, X., {Li}, W., {Filippenko}, A.~V., {et~al.} 2009{\natexlab{b}}, \apj,
  697, 380

\bibitem[{{Wang} {et~al.}(2008){Wang}, {Li}, {Filippenko}, {Krisciunas},
  {Suntzeff}, {Li}, {Zhang}, {Deng}, {Foley}, {Ganeshalingam}, {Li}, {Lou},
  {Qiu}, {Shang}, {Silverman}, {Zhang}, \& {Zhang}}]{wang2008}
{Wang}, X., {Li}, W., {Filippenko}, A.~V., {et~al.} 2008, \apj, 675, 626

\bibitem[{{Wang} {et~al.}(2013){Wang}, {Wang}, {Filippenko}, {Zhang}, \&
  {Zhao}}]{wang2013}
{Wang}, X., {Wang}, L., {Filippenko}, A.~V., {Zhang}, T., \& {Zhao}, X. 2013,
  Science, 340, 170

\bibitem[{{Webbink}(1984)}]{webbink1984}
{Webbink}, R.~F. 1984, \apj, 277, 355

\bibitem[{{Whelan} \& {Iben}(1973)}]{whelan1973}
{Whelan}, J. \& {Iben}, Icko, J. 1973, \apj, 186, 1007

\bibitem[{{Yang} {et~al.}(2022){Yang}, {Wang}, {Suntzeff}, {Hu}, {Aldoroty},
  {Brown}, {Krisciunas}, {Arcavi}, {Burke}, {Galbany}, {Hiramatsu},
  {Hosseinzadeh}, {Howell}, {McCully}, {Pellegrino}, \& {Valenti}}]{yang2022}
{Yang}, J., {Wang}, L., {Suntzeff}, N., {et~al.} 2022, \apj, 938, 83

\bibitem[{{Zhao} {et~al.}(2021){Zhao}, {Maeda}, {Wang}, \& {Sai}}]{zhao2021}
{Zhao}, X., {Maeda}, K., {Wang}, X., \& {Sai}, H. 2021, \mnras, 503, 4667

\end{thebibliography}

\begin{appendix}
\onecolumn

\section{Tables}

\begin{table}

\tiny
\centering
\caption{Classification and spectroscopic properties of SNe Ia from the volume limited ZTF DR2 sample.}
\label{tab:spec_p}
\begin{tabular}{l c c c c c c c c c } 

\hline\\[-0.5em]
ZTF Name & SN Name & Phase & Subtype & Wang Type & Branch Type & v$_{\text{\SiII}}$(6355) & pEW(6355) & v$_{\text{\SiII}}$(5972) & pEW(5972) \\

& & (d) & & & & (x10$^{\text{4}} $ km s$^{-1}$) & $(\angstrom)$ & (x10$^{\text{4}}$ km s$^{-1}$) & $(\angstrom)$\\[0.15em]

\hline\\[-0.8em]
\hline\\[-0.5em]

\multicolumn{9}{c}{\textit{d$_{DLR}$} $>$ 0.2}\\

\hline\\[-0.5em]

ZTF18aabtiph&2018yq&2.2&normal&NV&SS&1.18$\pm$$^{\text{0.02}}_{\text{0.02}}$&61.99$\pm$$^{\text{0.56}}_{\text{0.39}}$&1.16$\pm$$^{\text{0.02}}_{\text{0.02}}$&9.57$\pm$$^{\text{0.27}}_{\text{0.22}}$\\[0.15em]
ZTF18aahfgsk&2018aqk&2.3&normal&NV&CN&1.07$\pm$$^{\text{0.02}}_{\text{0.02}}$&79.23$\pm$$^{\text{1.44}}_{\text{1.83}}$&1.15$\pm$$^{\text{0.03}}_{\text{0.02}}$&15.31$\pm$$^{\text{1.35}}_{\text{1.79}}$\\[0.15em]
ZTF18aahfzea&2018aqh&$-$4.2&normal&NV&CN&1.10$\pm$$^{\text{0.02}}_{\text{0.02}}$&73.93$\pm$$^{\text{0.92}}_{\text{0.66}}$&1.10$\pm$$^{\text{0.02}}_{\text{0.02}}$&4.92$\pm$$^{\text{0.28}}_{\text{0.53}}$\\[0.15em]
ZTF18aahheaj&2018avp&2.7&normal&HV&BL&1.25$\pm$$^{\text{0.02}}_{\text{0.02}}$&109.65$\pm$$^{\text{2.4}}_{\text{1.25}}$&0.95$\pm$$^{\text{0.07}}_{\text{0.02}}$&13.68$\pm$$^{\text{1.73}}_{\text{4.90}}$\\[0.15em]
ZTF18aahjaxz&2018avg&$-$0.2&91bg-like&91bg&CL&1.04$\pm$$^{\text{0.02}}_{\text{0.02}}$&110.46$\pm$$^{\text{4.3}}_{\text{4.35}}$&1.06$\pm$$^{\text{0.02}}_{\text{0.02}}$&48.68$\pm$$^{\text{1.07}}_{\text{0.85}}$\\[0.15em]
ZTF18aahtjsc&2019cdd&$-$2.3&Ia-unclear&$-$&SS&0.96$\pm$$^{\text{0.02}}_{\text{0.02}}$&28.41$\pm$$^{\text{1.05}}_{\text{0.52}}$&$-$&10.88$^{*}$\\[0.15em]
ZTF18aaimxdx&2018bay&1.6&91bg-like&91bg&CL&1.05$\pm$$^{\text{0.06}}_{\text{0.06}}$&107.04$\pm$$^{\text{2.3}}_{\text{1.36}}$&0.94$\pm$$^{\text{0.06}}_{\text{0.06}}$&39.12$\pm$$^{\text{2.48}}_{\text{2.64}}$\\[0.15em]
ZTF18aajivpr&2020gaa&0.5&normal&NV&CN&1.06$\pm$$^{\text{0.02}}_{\text{0.02}}$&79.88$\pm$$^{\text{0.51}}_{\text{0.53}}$&1.02$\pm$$^{\text{0.02}}_{\text{0.02}}$&10.64$\pm$$^{\text{0.86}}_{\text{0.87}}$\\[0.15em]
ZTF18aajtlbf&2018bbz&$-$0.1&86G-like&$-$&CL&1.03$\pm$$^{\text{0.02}}_{\text{0.02}}$&122.62$\pm$$^{\text{1.3}}_{\text{2.33}}$&0.92$\pm$$^{\text{0.02}}_{\text{0.02}}$&46.88$\pm$$^{\text{2.80}}_{\text{1.41}}$\\[0.15em]
ZTF18aakglgw&2018aoy&$-$0.2&normal&HV&BL&1.33$\pm$$^{\text{0.02}}_{\text{0.02}}$&118.66$\pm$$^{\text{1.0}}_{\text{0.70}}$&1.13$\pm$$^{\text{0.02}}_{\text{0.02}}$&7.92$\pm$$^{\text{0.87}}_{\text{0.54}}$\\[0.15em]
ZTF18aaklpdo&2018hvf&$-$2.6&normal&NV&SS&1.13$\pm$$^{\text{0.02}}_{\text{0.02}}$&67.16$\pm$$^{\text{1.16}}_{\text{1.05}}$&1.08$\pm$$^{\text{0.03}}_{\text{0.03}}$&11.97$\pm$$^{\text{1.11}}_{\text{1.34}}$\\[0.15em]
ZTF18aalurka&2018awj&$-$3.9&normal&NV&CN&1.01$\pm$$^{\text{0.02}}_{\text{0.02}}$&104.26$\pm$$^{\text{4.4}}_{\text{1.91}}$&0.88$\pm$$^{\text{0.02}}_{\text{0.02}}$&27.09$\pm$$^{\text{1.06}}_{\text{0.59}}$\\[0.15em]
ZTF18aamlhee&2018zs&1.2&normal&NV&CN&1.06$\pm$$^{\text{0.02}}_{\text{0.02}}$&87.75$\pm$$^{\text{0.88}}_{\text{0.39}}$&0.96$\pm$$^{\text{0.02}}_{\text{0.02}}$&21.58$\pm$$^{\text{0.57}}_{\text{0.01}}$\\[0.15em]
ZTF18aaodrgt&2018bfd&1.5&normal&NV&BL&1.15$\pm$$^{\text{0.02}}_{\text{0.02}}$&125.11$\pm$$^{\text{0.5}}_{\text{0.54}}$&1.04$\pm$$^{\text{0.02}}_{\text{0.02}}$&25.19$\pm$$^{\text{0.99}}_{\text{1.05}}$\\[0.15em]
ZTF18aaoufjt&2019fyi&1.4&99aa-like&$-$&SS&1.36$\pm$$^{\text{0.02}}_{\text{0.02}}$&31.50$\pm$$^{\text{0.70}}_{\text{0.48}}$&$-$&6.15$^{*}$\\[0.15em]
ZTF18aaqfziz&2018bhp&4.2&normal&NV&BL&1.09$\pm$$^{\text{0.02}}_{\text{0.02}}$&122.68$\pm$$^{\text{2.0}}_{\text{1.94}}$&1.03$\pm$$^{\text{0.02}}_{\text{0.02}}$&24.07$\pm$$^{\text{0.66}}_{\text{1.25}}$\\[0.15em]
ZTF18aaroihe&2018bio&$-$0.3&04gs-like&$-$&CL&1.15$\pm$$^{\text{0.02}}_{\text{0.02}}$&96.61$\pm$$^{\text{2.30}}_{\text{2.07}}$&1.10$\pm$$^{\text{0.02}}_{\text{0.02}}$&30.55$\pm$$^{\text{2.75}}_{\text{2.97}}$\\[0.15em]
ZTF18aarymnz&2018azu&0.1&normal&HV&CN&1.24$\pm$$^{\text{0.02}}_{\text{0.02}}$&94.59$\pm$$^{\text{1.84}}_{\text{0.81}}$&1.22$\pm$$^{\text{0.02}}_{\text{0.02}}$&4.88$\pm$$^{\text{0.08}}_{\text{0.31}}$\\[0.15em]
ZTF18aaslhxt&2018btk&$-$2.3&normal&NV&CN&1.06$\pm$$^{\text{0.02}}_{\text{0.02}}$&99.43$\pm$$^{\text{1.97}}_{\text{2.55}}$&0.96$\pm$$^{\text{0.02}}_{\text{0.02}}$&15.13$\pm$$^{\text{0.72}}_{\text{1.32}}$\\[0.15em]
ZTF18aasprui&2018euz&0.4&04gs-like&$-$&CL&1.02$\pm$$^{\text{0.02}}_{\text{0.02}}$&107.13$\pm$$^{\text{0.4}}_{\text{0.42}}$&0.97$\pm$$^{\text{0.02}}_{\text{0.02}}$&34.20$\pm$$^{\text{0.45}}_{\text{0.19}}$\\[0.15em]

\multicolumn{9}{c}{.}\\[-0.9em]
\multicolumn{9}{c}{.}\\[-0.9em]
\multicolumn{9}{c}{.}\\[0.3em]

\hline\\[-0.8em]

\multicolumn{9}{c}{\textit{d$_{DLR}$}$^{a}$ $\leq$ 0.2}\\

\hline\\[-0.5em]

ZTF18aagstdc&	2018apn&	3.4&	normal&	NV&	CN&	1.19$\pm$$^{\text{0.06}}_{\text{0.06}}$&	98.35$\pm$$^{\text{0.74}}_{\text{0.55}}$&	1.20$\pm$$^{\text{0.06}}_{\text{0.06}}$&	18.69$\pm$$^{\text{0.28}}_{\text{0.31}}$\\[0.15em]
ZTF18aagtcxj&	2018aqm&	3.9&	normal&	NV&	SS&	1.09$\pm$$^{\text{0.02}}_{\text{0.02}}$&	69.86$\pm$$^{\text{1.26}}_{\text{1.08}}$&	1.10$\pm$$^{\text{0.02}}_{\text{0.02}}$&	18.13$\pm$$^{\text{0.67}}_{\text{0.72}}$\\[0.15em]
ZTF18aahfbqp&	2020acua&	-1.8&	Ia-unclear&	$-$&	SS&	0.99$\pm$$^{\text{0.02}}_{\text{0.02}}$&	29.94$\pm$$^{\text{1.02}}_{\text{0.63}}$&	$-$&	2.83$^{*}$\\[0.15em]
ZTF18aahjafd&	2018loz&	-1.5&	normal&	HV&	CN&	1.28$\pm$$^{\text{0.02}}_{\text{0.02}}$&	88.77$\pm$$^{\text{2.86}}_{\text{1.01}}$&	1.30$\pm$$^{\text{0.04}}_{\text{0.05}}$&	9.98$\pm$$^{\text{1.48}}_{\text{0.86}}$\\[0.15em]
ZTF18aahshhp&	2020dln&	-0.3&	Ia-unclear&	$-$&	SS&	0.83$\pm$$^{\text{0.02}}_{\text{0.02}}$&	14.23$\pm$$^{\text{0.49}}_{\text{0.65}}$&	$-$&	3.43$^{*}$\\[0.15em]
ZTF18aaifxvz&	2020lyo&	4.8&	Ia-unclear&	$-$&	SS&	1.05$\pm$$^{\text{0.02}}_{\text{0.02}}$&	48.24$\pm$$^{\text{0.96}}_{\text{0.75}}$&	1.00$\pm$$^{\text{0.02}}_{\text{0.02}}$&	4.03$\pm$$^{\text{0.33}}_{\text{0.47}}$\\[0.15em]
ZTF18aansqun&	2018dyp&	-1.9&	normal&	NV&	CN&	1.08$\pm$$^{\text{0.02}}_{\text{0.02}}$&	79.03$\pm$$^{\text{1.08}}_{\text{0.54}}$&	0.95$\pm$$^{\text{0.02}}_{\text{0.02}}$&	10.92$\pm$$^{\text{1.00}}_{\text{0.40}}$\\[0.15em]
ZTF18aaoxrup&	2020hdw&	-1.3&	04gs-like&	$-$&	CL&	1.09$\pm$$^{\text{0.06}}_{\text{0.06}}$&	93.76$\pm$$^{\text{0.61}}_{\text{0.41}}$&	0.96$\pm$$^{\text{0.06}}_{\text{0.06}}$&	30.76$\pm$$^{\text{1.10}}_{\text{0.81}}$\\[0.15em]
ZTF18aapqwyv&	2018bhc&	4.5&	Ia-unclear&	$-$&	SS&	1.02$\pm$$^{\text{0.04}}_{\text{0.04}}$&	54.39$\pm$$^{\text{3.07}}_{\text{3.01}}$&	1.10$\pm$$^{\text{0.06}}_{\text{0.06}}$&	14.02$\pm$$^{\text{2.62}}_{\text{2.97}}$\\[0.15em]
ZTF18aaqqhis&	2018ccy&	-1.6&	Ia-unclear&	$-$&	SS&	1.13$\pm$$^{\text{0.02}}_{\text{0.02}}$&	15.04$\pm$$^{\text{0.30}}_{\text{0.50}}$&	$-$&	3.91$^{*}$\\[0.15em]
ZTF18aarbaba&	2020mnu&	1.4&	normal&	NV&	CN&	1.15$\pm$$^{\text{0.02}}_{\text{0.02}}$&	98.25$\pm$$^{\text{1.85}}_{\text{0.75}}$&	$-$&	5.97$^{*}$\\[0.15em]
ZTF18aarcypa&	2018bil&	1.5&	Ia-unclear&	$-$&	CN&	0.96$\pm$$^{\text{0.02}}_{\text{0.02}}$&	78.21$\pm$$^{\text{1.95}}_{\text{0.35}}$&	0.82$\pm$$^{\text{0.02}}_{\text{0.02}}$&	22.31$\pm$$^{\text{0.52}}_{\text{0.10}}$\\[0.15em]
ZTF18aarikzk&	2020hll&	-1.3&	99aa-like&	$-$&	SS&	1.08$\pm$$^{\text{0.02}}_{\text{0.02}}$&	49.92$\pm$$^{\text{0.99}}_{\text{0.37}}$&	$-$&	19.77$^{*}$\\[0.15em]
ZTF18aaupmks&	2018bsr&	-4.8&	99aa-like&	$-$&	SS&	1.24$\pm$$^{\text{0.02}}_{\text{0.02}}$&	51.18$\pm$$^{\text{1.37}}_{\text{0.81}}$&	$-$&	7.18$^{*}$\\[0.15em]
ZTF18aavsilo&	2020jje&	3.9&	Ia-unclear&	$-$&	SS&	1.11$\pm$$^{\text{0.02}}_{\text{0.02}}$&	18.78$\pm$$^{\text{0.39}}_{\text{0.46}}$&	$-$&	5.77$^{*}$\\[0.15em]
ZTF18aawmxdx&	2018jti&	-3.4&	normal&	HV&	CN&	1.29$\pm$$^{\text{0.02}}_{\text{0.02}}$&	91.93$\pm$$^{\text{0.64}}_{\text{0.32}}$&	1.23$\pm$$^{\text{0.02}}_{\text{0.02}}$&	9.11$\pm$$^{\text{0.58}}_{\text{0.95}}$\\[0.15em]
ZTF18abbikrz&	2020pwn&	-2.0&	Ia-unclear&	$-$&	SS&	0.94$\pm$$^{\text{0.02}}_{\text{0.02}}$&	33.62$\pm$$^{\text{1.92}}_{\text{1.50}}$&	$-$&	20.32$^{*}$\\[0.15em]
ZTF18abespgb&	2018dyg&	-4.0&	normal&	NV&	CN&	1.16$\pm$$^{\text{0.06}}_{\text{0.06}}$&	83.29$\pm$$^{\text{1.72}}_{\text{0.82}}$&	1.16$\pm$$^{\text{0.06}}_{\text{0.06}}$&	5.82$\pm$$^{\text{0.53}}_{\text{0.10}}$\\[0.15em]
ZTF18abhbqbh&	2020kwk&	3.5&	Ia-unclear&	$-$&	SS&	1.09$\pm$$^{\text{0.02}}_{\text{0.02}}$&	35.38$\pm$$^{\text{1.28}}_{\text{0.94}}$&	1.24$\pm$$^{\text{0.02}}_{\text{0.03}}$&	7.92$\pm$$^{\text{0.82}}_{\text{0.36}}$\\[0.15em]
ZTF18abjhack&	2018crv&	0.5&	normal&	NV&	CN&	1.08$\pm$$^{\text{0.06}}_{\text{0.06}}$&	92.67$\pm$$^{\text{1.05}}_{\text{0.45}}$&	$-$&	5.33$^{*}$\\[0.15em]

\multicolumn{9}{c}{.}\\[-0.9em]
\multicolumn{9}{c}{.}\\[-0.9em]
\multicolumn{9}{c}{.}\\[0.3em]

\hline\\[-0.5em]
\end{tabular}

{\raggedright($^a$) In this study only SNe Ia with a value of \textit{d$_{DLR}$} higher than 0.2 is used. However the results from lower \textit{d$_{DLR}$} values are also listed here.\par}
{\raggedright($^*$) estimated pEW values for \SiIIf\ features as upper limits.\par}
{\raggedright The complete table is accessible within the online journal. Here, a section is displayed to offer direction on its structure and information included.\par}

\end{table}

\end{appendix}

\end{document}